\documentclass[transmag]{IEEEtran}

\IEEEoverridecommandlockouts
\usepackage{graphicx}
\usepackage{soul}

\ifCLASSOPTIONcompsoc
    \usepackage[caption=false, font=normalsize, labelfont=sf, textfont=sf]{subfig}
\else
\usepackage[caption=false, font=footnotesize]{subfig}
\fi
\usepackage{url}
\usepackage{mdwmath}     
\usepackage[table]{xcolor}
\usepackage{amsfonts,amsmath,amssymb,amsxtra,amsbsy,floatflt,algorithm,algpseudocode}
\usepackage{indentfirst}
\usepackage[T1, OT1]{fontenc}
\DeclareTextSymbolDefault{\DH}{T1}
\usepackage{tabularx}
\usepackage{diagbox}
\usepackage{booktabs}
\usepackage{multirow}
\usepackage{nth}
\usepackage{comment}
\usepackage{epstopdf}
\usepackage{bbold}
\usepackage{multirow}
\usepackage{hhline}
\usepackage{siunitx}
\usepackage[sort,noadjust]{cite}
\usepackage{booktabs}
\usepackage{caption}
\usepackage{graphicx}
\usepackage[shortlabels]{enumitem}

\newtheorem{problem}{Problem}
\newcommand{\vv}[1]{\boldsymbol{#1}}

\newcommand{\change}[1]{{\color{black}{#1}}}

\newcommand{\tr}{\mathrm{tr}}
\newcommand{\Real}{\mbox{$\mathbb{R}$}}
\newcommand{\Compl}{\mbox{$\mathbb{C}$}}
\newcommand{\tran}{\mathrm{T}}
\newcommand{\herm}{\mathrm{H}}
\newcommand{\Exp}{\mathbb{E}}
\newcommand{\rmmult}{\scriptscriptstyle\mathrm{mult}}

\newcommand{\name}{RiFe}
\def\BibTeX{{\rm B\kern-.05em{\sc i\kern-.025em b}\kern-.08em T\kern-.1667em\lower.7ex\hbox{E}\kern-.125emX}}

\begin{document}

\title{RISe of Flight: RIS-Empowered UAV Communications \\for Robust and Reliable Air-to-Ground Networks}
\author{Placido Mursia, \IEEEmembership{Student Member, IEEE}, Francesco Devoti, \IEEEmembership{Member, IEEE},\\ Vincenzo Sciancalepore, \IEEEmembership{Senior Member, IEEE}, and Xavier Costa-Perez,
\IEEEmembership{Senior Member, IEEE}
\thanks{Manuscript received xxx XX, xxx; revised XXX xx, xx; accepted XXX x, xxx. 

The work was supported by the European Union H-2020 Project RISE-6G under grant 101017011.
}
\thanks{P. Mursia, F. Devoti and V. Sciancalepore are with NEC Laboratories Europe GmbH, Heidelberg, Germany (e-mails: \{name.surname\}@neclab.eu).}
\thanks{X. Costa-Perez is with NEC Laboratories Europe GmbH, Heidelberg, Germany, and i2CAT Foundation and ICREA, Barcelona, Spain (e-mail: xavier.costa@neclab.eu).}}

\IEEEtitleabstractindextext{\begin{abstract}
Next generation mobile networks need to expand towards uncharted territories in order to enable the digital transformation of society.
In this context, aerial devices such as unmanned aerial vehicles (UAVs) 
are expected to address this gap 
in hard-to-reach locations. However, limited battery-life is an obstacle for the successful spread of such solutions. 
Reconfigurable intelligent surfaces (RISs) represent a promising solution addressing this challenge since on-board passive and lightweight controllable devices can efficiently reflect the signal propagation from the ground BSs towards specific target areas. 
In this paper, we focus on air-to-ground networks where UAVs equipped with RIS can fly over selected areas to provide connectivity. 
In particular, we study how to optimally compensate \emph{flight effects} and propose \emph{\name{}} as well as its practical implementation \emph{Fair-\name{}} that automatically configure RIS parameters accounting for undesired UAV oscillations due to adverse atmospheric conditions. Our results show that both algorithms provide robustness and reliability while outperforming state-of-the-art solutions in the multiple conditions studied.

\end{abstract}

}

\maketitle

\section{Introduction}
Unmanned aerial vehicles (UAVs)\,\footnote{The terms UAV and drone are used interchangeably throughout the paper.} are increasingly becoming part of our lives by enhancing how we work, e.g., package delivery, how we entertain ourselves and how we extend the safety and security of our society. Li-ion batteries have already reached a mature state thereby allowing limited-size devices to fly towards hard-to-reach locations in very short time~\cite{Moz17,Gup16}. Recently, the telecom industry and academia made a big effort in bringing flexibility and agility to advanced wireless networks that rely on flying access points~\cite{Moz18}, namely air-to-ground networks. UAVs have demonstrated to be suitable for readily establishing a line-of-sight (LoS) link towards ground users thereby avoiding obstacles that impair the overall communication quality~\cite{Fen19, Zen19}. Dense-urban scenarios  further exacerbate the obstruction issues making UAVs a viable solution for building reliable networks on-demand.

\begin{figure}[t]
    \centering
    \includegraphics[width=\linewidth]{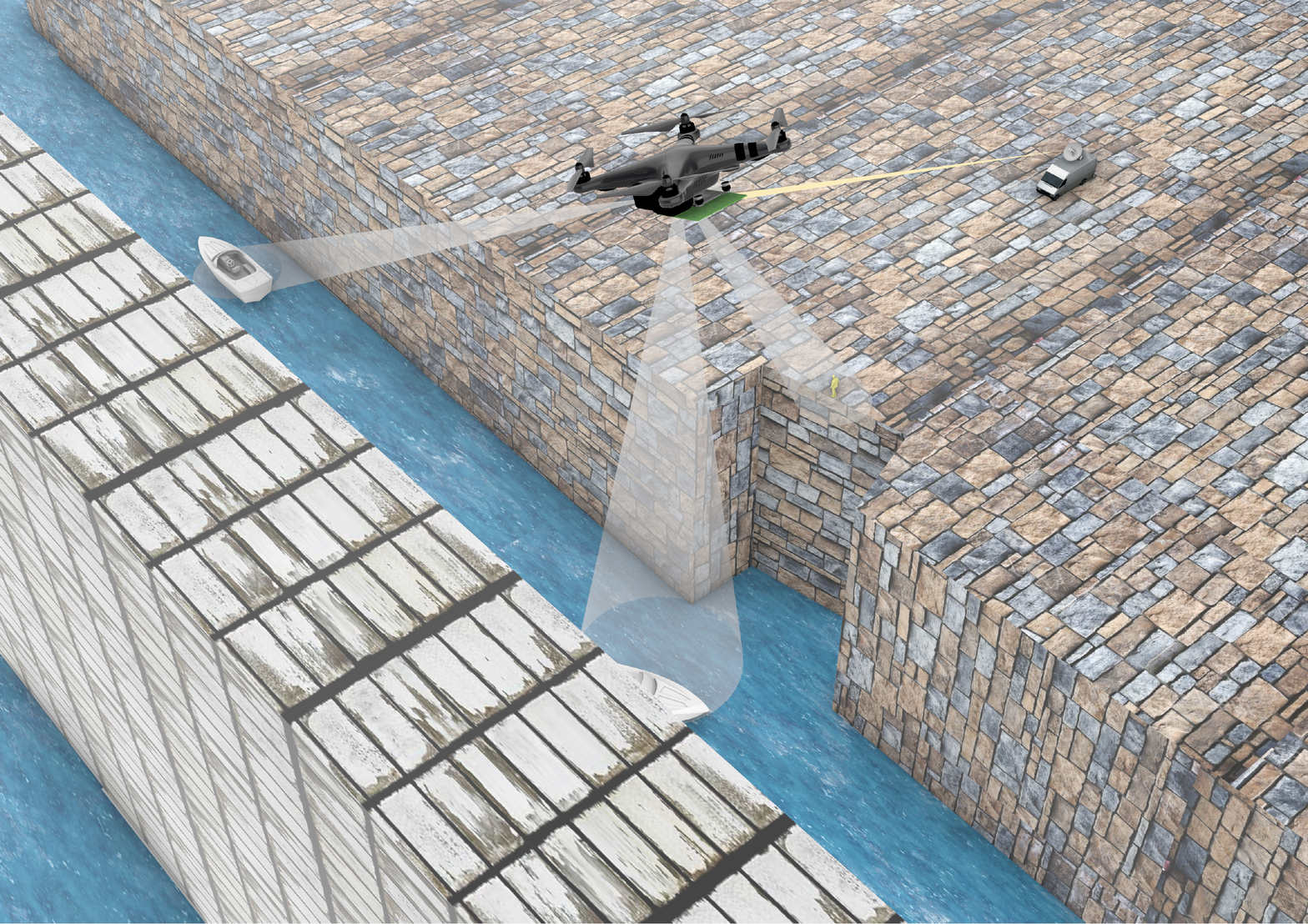}
    \caption{UAV equipped with a lightweight RIS covering first-responder team and victims in an emergency scenario.}
    \label{fig:scenario}
\end{figure}

Due to their fast-deployment property, UAVs have been identified as a key technology to deal with emergency situations so as to expedite rescue operations, e.g., assisting first-responder teams, in areas where there might not be connectivity or when the network infrastructure is temporarily damaged or unavailable~\cite{Pan19}. Interestingly, UAVs are proposed to bring back-up connectivity in such areas and/or leverage on advanced sensing and localization techniques exploiting the cellular protocol stack to find missing people~\cite{Alb21}. However, since they are envisioned as flying mobile base stations (BSs) carrying one or more active antennas, a significant increase of the total power consumption and, in turn, battery drain issues are expected due to $i$) the  weight of active elements, $ii$) the power irradiated to reach ground targets and $iii$) the additional power burden required to establish a backhaul link and process incoming packets.

To overcome the above-mentioned issues, lightweight and low-energy equipment is needed on board. In this context, reconfigurable intelligent surfaces (RISs) are currently gaining much attention owing to their ability to control the propagation environment by altering reflection, absorption and amplitude properties of the material where the signal bounces off~\cite{Hu18, Bas19, Mur20, Cal21}. This introduces a new powerful tool that allows to efficiently redirect the signal propagation at very limited power expense: varactor diodes can apply a phase shift and/or absorb the incoming signal in a real-time reconfigurable way. 
Passive, flexible and configurable elements match a variety of envisioned applications spanning among electromagnetic field exposure efficiency (EMFEE)-boosted communications, extreme-accurate localization mechanisms and user-centric quality-of-service-enhanced connections~\cite{Zha20}. 



As depicted in Fig.~\ref{fig:scenario}, conventional UAV-based systems may be combined with comfortably sized RISs to effectively reflect incoming signal towards hard-to-reach locations while keeping the power consumption very low~\cite{Alf20, Abd20}. While appropriate RIS modeling assumes the location of transmitters, receivers and RIS to be known, the high mobility property of UAVs introduces unprecedented random variables in such a complex analysis. \change{Interestingly, \cite{zhang21_uav} discusses design challenges of UAV-mounted RIS-assisted terrestrial communication. Indeed, the overall communication performance may be negatively affected when small variations of the orientation of the RIS and/or its position occur. Sophisticated sensors system and GPS antennas are built to keep the drone stable and its location fixed while hovering on selected areas: when adverse meteorological conditions strike, drone movement counteractions are automatically taken but still orientation oscillations or location perturbation may result in an instantaneously-wrong RIS configuration. }


In this paper, we focus on the above-mentioned scenario where UAVs are provided with passive RISs to support first responder teams in selected target areas. 
\change{Our solution, namely \emph{\name{}}, takes into account the statistical properties of unwanted UAV oscillations due to meteorological phenomena while optimizing RIS parameters such as phase shift and reflection coefficient. To the best of our knowledge, \emph{\name{}} is the first of its kind tackling such undesired oscillations while steering signal reflections to build a robust and reliable air-to-ground network solution.}

\subsection{Contributions}

The main novelty of this paper stems from the design of an optimization framework that addresses undesired UAV oscillations---caused by meteorological phenomena---which exacerbates when high altitudes are considered. This is in marked contrast with prior works that 
usually assume a perfectly stable flight. 
Since UAV flight perturbations are always present and inevitable in real-life situations, our solution \name{}, aims at \emph{maximizing the worst SNR averaged over the unwanted perturbations on the UAV in a target area wherein a number of receivers is present}. \name{} is based solely on second-order statistics of said perturbations and of the position of the receivers. Thus, an advantage of such approach is the reduction in the overhead necessary to acquire instantaneous channel information. 

In addition to the optimization framework, we extend \name{} to account for practical considerations such as the need to update the RIS parameters due to rapidly-changing channel statistics, the mobility of the UAV as well as  complexity issues and rename it as  Fair-\name{}. Our numerical results show, as expected, that our approach outperforms state-of-the-art solutions which assume perfectly stable flights. Indeed, our solution results in a considerably higher received SNR at target locations due to more robust passive beamforming at the RIS. The main four contributions of this paper are thus summarized in the following.
\begin{enumerate}[start=1,label={\bfseries C\arabic*:}]
    \item A novel mathematical framework to maximize the worst SNR within a given target area by taking into account unwanted rotations on the position of the surface of the RIS due to flight perturbations.
    \item A low-complexity approach \name{} that tackles the solution of the aforementioned problem by designing both the precoding vector at the BS and the adequate RIS configuration.
    \item An easy-to-develop solution Fair-\name{} that accounts for practical considerations such as time and/or complexity constraints on the optimization routine and re-configuration of the RIS.
    \item Exhaustive numerical results that show substantial gains in terms of received average SNR at the target locations with respect to state-of-the-art solutions.
\end{enumerate}

\change{The remainder of this paper is structured as follows. In Section~\ref{sec:Related_work} we review the state of the art when integrating RISs in UAV-based communications and highlight the differences with respect to our proposed approach. Section~\ref{sec:sys_model} introduces the system model and defines the key physical quantities of interest. In Section~\ref{sec:RiFe} we give the problem formulation to optimize the considered metric and propose \name{} to tackle its solution. In Section~\ref{sec:practical_cons} we propose Fair-\name{}, which takes into account practical considerations such as timing constraints and UAV mobility. Section~\ref{sec:Perf_eval} presents numerical results to evaluate the performance of the proposed algorithms. Lastly, Section~\ref{sec:concl} concludes the paper.}

\subsection{Notation}

We use italic letters to denote scalars, whereas vectors and matrices are denoted by bold-face lower-case and upper-case letters, respectively. We let $\Compl$ and $\Real$ denote the set of complex and real numbers, respectively. We use $\Compl^N$ and $\Compl^{N\times M}$ to represent the sets of $N$-dimensional complex vectors and  $N\times M$ complex matrices, respectively. $\vv{I}_N$ stands for the identity matrix of size $N$ while $j$ represents the imaginary index. Vectors are denoted by default as column vectors while subscripts represent an element in a vector. For instance, $\vv{x} =   [x_1, \dots,  x_n]^{\tran}$ is a vector from $\Compl^N$ and  $x_i$ is its $i$-th component. The symbols $(\cdot)^{\tran}$, $\Exp[\cdot]$, $(\cdot)^*$, $\tr(\cdot)$ and $(\cdot)^{\herm}$ represent the transpose, expectation, complex conjugate, trace, and Hermitian transpose operator, respectively. $\|\cdot\|$ denotes the L2-norm of a vector. Lastly, $|\cdot|$ and $\angle\cdot $ denote the absolute value and angle of a complex number, respectively.

\section{Related work}\label{sec:Related_work}

Air-to-ground (A2G) communication has been recently promoted as an upcoming technology to deliver advanced and sophisticated services. In~\cite{jiang_tcom20}, the authors introduce effective A2G channel models while empirically validating them via realistic measurements. In particular, A2G can be implemented to support vehicular-based use-cases~\cite{liu_let21} or to assist emergency communication~\cite{lin_lett19}. In both, the authors propose an optimization-based approach to maximize the end-user coverage during the UAV flight. 

Recently, an integration of RIS in UAV-based communications has been proposed as a means to enhance the performance of air-to-ground networks (see, e.g., \cite{DiR20,Wu20} and references therein). In \cite{Ma19}, the authors propose to use a RIS to increase the power of the received signal at the UAV. However, this approach has the limitation of requiring both the UAV and the user to lie on the same side with respect to the RIS~\cite{Wu21}. 

Hence, a more feasible solution is to install a lightweight RIS on board the UAV. In this regard, several works have investigated the joint UAV trajectory and RIS parameters optimization problem~\cite{Zha19,Lon20,Li20,Lu20,Lu20_2}. In \cite{Zha19} the authors propose to employ a reinforcement learning approach based on Q-learning and neural networks to optimize both the trajectory of the UAV so as to maintain a LoS link with the ground user, and the RIS parameters to maximize the downlink transmission capacity. In~\cite{Lon20} the authors tackle the maximization of the secure energy efficiency of the system via joint optimization of the trajectory of the UAV, the phase shifts at the RIS, the user association and transmit power by means of successive convex approximation. 

A similar technique is also used in \cite{Li20} to maximize the achievable rate of a single ground user. In \cite{Lu20,Lu20_2} the authors consider the problem of maximizing the worst SNR in a given target area to be covered by the BS by proper choice of the RIS configuration and placement of the aerial platform. The resulting non-convex problem is solved by decoupling the optimization in maximizing the worst-case array gain to find the RIS configuration, and balancing the resulting angular span and path loss of the equivalent downlink channel to find the placement of the aerial platform.

However, existing works assume for simplicity that aerial platforms (e.g., UAV or HAP) has a perfectly stable flight. Inspired by real-life scenarios, in this paper we consider the case where the UAV is subject to perturbations due to meteorological phenomena. Specifically, we pioneer a mathematical framework that takes into account undesired perturbations while continuously configuring the on-board RIS to pursue the overall performance maximization in terms of fairness within a target area and minimum experienced SNR.

\section{System model}\label{sec:sys_model}

 \begin{figure}[t]
        \center
        \includegraphics[width=\linewidth]{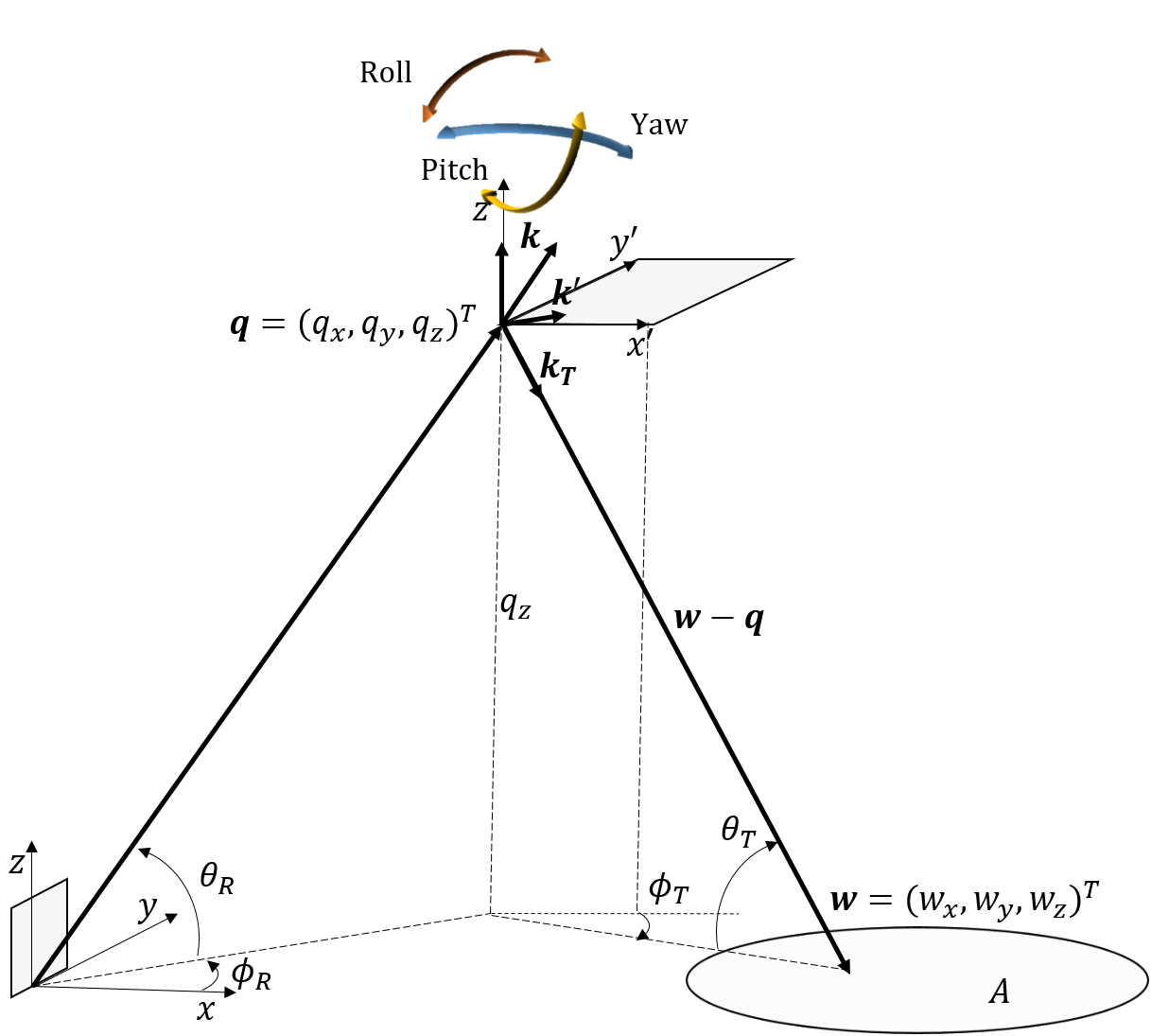}
        \caption{\change{\label{fig:geometry} Geometrical representation of the considered scenario including the transmitter (bottom-left side), the RIS (top part) and the target area $A$ (bottom-right side).}}
\end{figure}
      
We describe hereafter the adopted system model whose geometrical representation is depicted in Fig.~\ref{fig:geometry}. We consider a base station (BS), or general transmitter, located at the origin and equipped with $M$ antennas whose signal shall cover a target area $\mathcal{A}$ wherein first responders and/or victims are present. Moreover, we consider a RIS mounted on a UAV, composed of $N$ reflecting elements and reflecting the signal coming from the BS towards the target area. Specifically, each aforementioned receiver shall be reached by a signal experiencing a minimum SNR to successfully decode the upcoming packets. We assume that the position of the first responder team is known with some uncertainty, whereas only a probability distribution function (pdf) of the spatial position of the victims is known. 

Considering that the UAV is hovering, it might be subject to perturbations due to the wind or other meteorological phenomena, which result in undesired roll, yaw and pitch of the surface of the RIS. Let $\vv{q}=[q_x,q_y,q_z]^{\tran}$ denote the position of the center of the RIS and
$\phi_R$ and $\theta_R$ denote the azimuth and elevation angles of the geographical path connecting the BS to the RIS, respectively. Hence, the coordinates of $\vv{q}$ can be expressed as $q_x = \|\vv{q}\| \cos(\theta_R) \cos(\phi_R)$, $q_y = \|\vv{q}\|\cos(\theta_R)\sin(\phi_R)$, and $q_z = \|\vv{q}\|\sin(\theta_R)$, respectively. Furthermore, let $\vv{r}=[\psi_x,\psi_y,\psi_z]^{\tran}$, which takes into account the possible perturbations on the orientation of the surface of the RIS, where the random variables (RVs) $\psi_x$, $\psi_y$, and $\psi_z$ represent the rotations on the $x$, $y$, and $z$ axes, respectively. 

We assume that such rotations are mutually independent and normally distributed with zero mean and variance $\sigma^2_{\psi,x}$, $\sigma^2_{\psi,y}$, and $\sigma^2_{\psi,z}$, respectively. We further assume that the BS has the perfect knowledge of the statistics of $\vv{r}$. Let $\vv{w}=[w_x,w_y, w_z]^{\tran} \in \mathcal{A}$---distributed according to a given probability density function expressed as per $f(\vv{w})$---be a point in the target area to be covered and let $\phi_T$ and $\theta_T$ denote the corresponding azimuth and elevation angles with respect to the RIS, respectively. The coordinates of $\vv{w}$ can thus be expressed as $w_x = \|\vv{w}\| \cos(\theta_T)\cos(\phi_T)$, \change{$w_y = \|\vv{w}\| \cos(\theta_T)\sin(\phi_T)$,} and $w_z = \|\vv{w}\| \sin(\theta_T)$, respectively. Note that in the case of first responder team, we model $f_{\vv{w}}(\vv{w})$ as the union of two-dimensional Gaussian distributions whose means correspond to the nominal positions of each first responder member and with given variances along each direction. Conversely, when considering victims $f_{\vv{w}}(\vv{w})$ can be modeled again as a two-dimensional Gaussian distribution with mean equal to the center of the target area and with given variances along the two directions in space. If both first responder team and victims are present then $f_{\vv{w}}(\vv{w})$ is modeled as the union of the two aforementioned cases.

The distance between the source node and the RIS is defined as the following
\begin{align}
d_1(\vv{q}) \triangleq \|\vv{q}\|,   
\end{align}
whereas the distance between the RIS and the target position $\vv{w}$ is defined as the following
\begin{align}
d_2(\vv{q},\vv{w})\triangleq\|\vv{q} - \vv{w}\|.
\end{align}
Hence, the overall length of the communication path is equal to $d(\vv{q},\vv{w})=d_1(\vv{q})+d_2(\vv{q},\vv{w})$.

\change{Given the position of the RIS with respect to the BS and the target area $\mathcal{A}$ and the fact that we assume the drone to be maneuvered by first responder teams, the communication link between the BS and the RIS is assumed to be in line-of-sight (LoS) condition with a very high probability (see, e.g., \cite{Abd20,Alf20_2,Li20}).}
Therefore, the channel power gain from the source to the RIS is defined as per
\begin{equation}
    \beta_1(\vv{q}) \triangleq \frac{\beta_0}{d_1^{2}(\vv{q})}  = \frac{\beta_0}{\|\vv{q}\|^2},  
\end{equation}
where $\beta_0$ is the channel power gain at a reference distance. 
\change{In a similar way, by adopting the one-ring channel model (see, e.g., \cite{Shi00}) we define the channel power gain between the RIS and the target point $\vv{w}$ as the following
}
\begin{equation}
    \beta_2(\vv{q},\vv{w}) \triangleq \frac{\beta_0}{d_2^{2}(\vv{q},\vv{w})}  = \frac{\beta_0}{\|\vv{q} - \vv{w}\|^2}.
\end{equation}

We consider a relative reference system $\Omega$ with origin in the right bottom corner of the RIS and with the $(x,y)$-plane lying on the RIS surface as shown in Fig.~\ref{fig:geometry}. The position of the $i$-th element of the RIS with respect to $\Omega$ is defined by the vector $\vv{c}^{(\Omega)}_i=[x_{i},y_{i},z_{i}]^{\tran}$. The position of the $i$-th element of the RIS in the absolute coordinate system, which has the origin located at the BS, is thus expressed as 
\begin{align}
    \vv{c}_i = \vv{R}(\vv{r})\vv{c}^{(\Omega)}_i + \vv{q},    
\end{align}
where the rotation matrix $\vv{R}(\vv{r})\in\Real^{3\times 3}$ takes into account the possible aforementioned rotations of the RIS with respect to the absolute reference system and it is defined as the following

\begin{equation}
\vv{R}(\vv{r})\triangleq\vv{R}_x(\psi_x)\vv{R}_y(\psi_y)\vv{R}_z(\psi_z)
\end{equation}
with the rotation matrices on the $x$, $y$, and $z$ axes defined as
\begin{align}
 \vv{R}_x(\psi_x) & \triangleq \left(
\begin{matrix}
1 & 0 & 0 \\
0 & \cos{\psi_x} & -\sin{\psi_x}\\
0 & \sin{\psi_x} & \cos{\psi_x}\\
\end{matrix} \right),\\
\vv{R}_y(\psi_y) & \triangleq \left(
\begin{matrix}
\cos{\psi_y} & 0 & \sin{\psi_y}\\
0 & 1 & 0 \\
-\sin{\psi_y} &  0 & \cos{\psi_y}
\end{matrix} \right),
\end{align}
and
\begin{align}
\vv{R}_z(\psi_z) & \triangleq \left(
\begin{matrix}
\cos{\psi_z} & -\sin{\psi_z} & 0\\
\sin{\psi_z} & \cos{\psi_z} & 0 \\
0 & 0 & 1 
\end{matrix}\right),
\end{align}
respectively.





We assume that the drone is flying at an altitude such that the incoming signal can be considered as a plane wave\footnote{We consider the far-field beamforming case from the UAV equipped with a RIS and the ground. We refer the reader to~\cite{Tang21} for more details.} with propagation direction identified by $\vv{k} = \frac{\vv{q}}{\|\vv{q}\|}$. Hence, the phase shift on the $i$-th element of the RIS related to the incoming signal from the BS is defined as the following\footnote{Note that we are interested in the phase shift on the surface of the RIS, thus we can safely ignore the coordinate along the $z$ axis to reduce notation clutter.}
\begin{align}
    \phi_{i}^R(\vv{q},\vv{r}) &\triangleq \frac{2\pi}{\lambda}\vv{k}^{\tran}\vv{R}(\vv{r})\begin{bmatrix}
    \vv{I}_2 & \vv{0} \\ \vv{0}^{\tran} & 0
    \end{bmatrix}\vv{c}^{(\Omega)}_i \\
                   &= \frac{2\pi}{\lambda} \cos \theta_R
                   \begin{bmatrix}
                   \cos(\phi_R-\psi_z)\cos\psi_y \\
                   \sin(\phi_R-\psi_z)\cos\psi_x
                   \end{bmatrix}^{\tran} \begin{bmatrix}
                   x_i \\ y_i
                   \end{bmatrix} \nonumber\\
                   & \phantom{=} + \frac{2\pi}{\lambda}\sin\theta_R
                   \begin{bmatrix}
                   \sin\psi_y\\
                   \sin\psi_x
                   \end{bmatrix}^{\tran} \begin{bmatrix}
                   x_i \\ y_i
                   \end{bmatrix} \\
                   &=\vv{v}_R(\vv{q},\vv{r})^{\tran}\overline{\vv{c}}^{(\Omega)}_i,
\end{align}
where $\lambda$ represents the signal wavelength\change{, $\vv{0}$ is the zero-vector of size $2\times 1$,} and where we have defined $\overline{\vv{c}}^{(\Omega)}_i$, which is equal to the first two elements of $\vv{c}^{(\Omega)}_i$ and the vector $\vv{v}_R(\vv{q},\vv{r})$ that describes the spatial frequency on the surface of the RIS corresponding to the incoming signal from the BS. As expected, the phase shifts on the elements of the RIS depend on the undesired changes of the orientation of the UAV $\vv{r}$. The elements of the receiving array response vector $\vv{a}_R(\vv{q},\vv{r}) \in \mathbb{C}^{N\times1}$ at the RIS are thus expressed as
\begin{equation}
    [\vv{a}_R(\vv{q},\vv{r})]_i = e^{-j \, \phi_{i}^R(\vv{q},\vv{r})}, \ i = 1\dots N.
\end{equation}

In a similar way, the phase shift on the $i$-th element of the RIS and the corresponding value of the antenna array response when reflecting the signal towards the target point $\vv{w}$ are defined as the following
\begin{align}
    \phi_{i}^T(\vv{q},\vv{w},\vv{r}) &\triangleq\frac{2\pi}{\lambda}\cos{\theta_T}
                    \begin{bmatrix}
                    \cos{(\phi_T-\psi_z)}\cos{\psi_y}\\
                    \sin{(\phi_T-\psi_z)}\cos{\psi_x}
                    \end{bmatrix}^{\tran}\overline{\vv{c}}^{(\Omega)}_i\nonumber\\
                   &\phantom{=}+\frac{2\pi}{\lambda}\sin{\theta_T}
                   \begin{bmatrix}
                   \sin{\psi_y}\\
                   \sin{\psi_x}
                   \end{bmatrix}^{\tran}\overline{\vv{c}}^{(\Omega)}_i\\
                   &=\vv{v}_T(\vv{q},\vv{w},\vv{r})^{\tran}\overline{\vv{c}}^{(\Omega)}_i,
\end{align}
and
\begin{equation}
    [\vv{a}_T(\vv{q},\vv{w},\vv{r})]_i = e^{-j\,\phi^T_{i}(\vv{q},\vv{w},\vv{r})}, \ i = 1 \dots N,
\end{equation}
respectively, where we define the vector $\vv{v}_T(\vv{q},\vv{r})$ that describes the spatial frequency on the surface of the RIS when reflecting the signal towards the target area.

The random phase shifts $\phi^R_{i}$ and $\phi^T_{i}$ refer respectively to the incoming and departing wave with respect to the RIS surface. Interestingly, they need to be properly aligned by suitably adjusting the RIS parameters: the role of the RIS is to maximize the network performance while statistically counteracting the undesired rotations of the RIS surface $\vv{r}$ caused by perturbations.

We further define the channel between the BS and the RIS as the following
\begin{equation}
    \vv{G}(\vv{q},\vv{r}) \triangleq \sqrt{\beta_1(\vv{q})}\vv{a}_R(\vv{q},\vv{r})\vv{a}^{\herm}_{BS}(\vv{q}) \in \mathbb{C}^{N\times M},
    \label{eq:channel_bs_ris}
\end{equation}
where $\vv{a}_{BS}$ is the antenna array response at the BS. The obtained channel between the RIS and the target position $\vv{w}$ is given by
\begin{equation}
    \vv{h}(\vv{q},\vv{w},\vv{r}) \triangleq \sqrt{\beta_2(\vv{q},\vv{w})}\vv{a}_T(\vv{q},\vv{w},\vv{r}) \in \mathbb{C}^{N\times 1}.
    \label{eq:channel_ris_w}
\end{equation}

From Eqs.~\eqref{eq:channel_bs_ris} and \eqref{eq:channel_ris_w}, we can write the receive signal at the intended location $\vv{w}$ as the following
\begin{align}
    y(\vv{q},\vv{w},\vv{r},\vv{\Theta},\vv{v}) & = \vv{h}^{\herm}(\vv{q},\vv{w},\vv{r})\vv{\Theta}\vv{G}(\vv{q},\vv{r})\vv{v}s + n,
\end{align}
where $\vv{v}\in \Compl^{M}$ is the precoding vector at the BS,  $\vv{\Theta} = \mathrm{diag}(\alpha_1e^{j\theta_1},\ldots,\alpha_Ne^{j\theta_N})$ represents the phase shifts (to be optimized) at the RIS with $|\alpha_{\ell}|^2\leq 1$ and $\theta_{\ell} \in (0,2\pi]$, $\forall  \ell$. Furthermore, $s\in \Compl$ is the transmitted symbol with $\Exp[|s|^2]=1$ and $n\sim \mathcal{CN}(0,\sigma_n^2)$ is the noise term which is assumed to be independent from $s$. \change{Note that in the following we propose a method to optimize directly the RIS configuration in matrix $\vv{\Theta}$ and not the absorption coefficients $\{\alpha_{\ell}\}_{\ell=1}^N$ and the phase shifts $\{\theta_{\ell}\}_{\ell=1}^N$. However, the latter can be recovered by simply setting
\begin{align}
    \alpha_{\ell} &= |[\vv{\Theta}]_{\ell \ell}|,
\end{align}
and
\begin{align}
    \theta_{\ell} &= \angle[\vv{\Theta}]_{\ell \ell}\quad \forall \ell,
\end{align}
respectively.}

\section{RIS to compensate flight effects}\label{sec:RiFe}

Our objective is to guarantee coverage over a given target area $\mathcal{A}$, i.e., ensuring that each receiver obtains a sufficiently high SNR, by taking into account the undesired rotations of the surface of the RIS caused by perturbations on the UAV. In this regard, in the following we define the received SNR at a given target location $\vv{w}$ and formulate an effective optimization problem that pursues the worst SNR maximization among all possible target locations, accounting for possible perturbations. We therefore provide a solution to such problem based on semidefinite relaxation (SDR) and Monte Carlo sampling denoted as \emph{RIS to compensate flight effects (\name)}.

\subsection{Problem Formulation}\label{sec:prob_form}

Let us consider the received SNR at position $\vv{w}$, which is given by the following
\begin{align}
    \mathrm{SNR}(\vv{q},\vv{w},\vv{r},\vv{\Theta},\vv{v}) & = \frac{|\vv{h}^{\herm}(\vv{q},\vv{w},\vv{r})\vv{\Theta}\vv{G}(\vv{q},\vv{r})\vv{v}|^2}{\sigma_n^2} \\
    & = \frac{|\vv{a}_T^{\herm}(\vv{q},\vv{w},\vv{r})\vv{\Theta}\vv{a}_R(\vv{q},\vv{r})\vv{a}_{BS}^{\herm}(\vv{q})\vv{v}|^2}{\|\vv{q}\|^2\|\vv{q}-\vv{w}\|^2\sigma_n^2/\beta_0^2}.\label{eq:snr_2}
\end{align}
Our objective is to maximize the worst-case SNR within the target region via the proper choice of both beamforming vector at the BS and the RIS configuration while accounting for the perturbations, which are identified by means of $\vv{r}$, i.e.,
\begin{problem}[P\_1]\label{eq:Prob}
\begin{align}
     \displaystyle \max_{\vv{\Theta},\vv{v}} \ \min_{\substack{\vv{w}\in \mathcal{A} \\ \vv{w}\sim f_{\vv{w}}(\vv{w})}} &  \Exp[\mathrm{SNR}(\vv{q},\vv{w},\vv{r},\vv{\Theta},\vv{v})] \\
     \mathrm{s.t.} & \quad \|\vv{v}\|^2 \leq P; \\
     & \quad  [\vv{\Theta}]_{i \ell} = 0, \quad \forall i\neq \ell;\\
     & \quad  |[\vv{\Theta}]_{ii}|^2 \leq 1, \quad i = 1,\ldots,N;
\end{align}
\end{problem}
where $P$ is the power budget available at the BS and the expectation is calculated over the RV $\vv{r}$. Note that the constraints on the matrix $\vv{\Theta}$ ensure that the RIS remains a passive structure without amplifying the incoming signal. Problem~\ref{eq:Prob} is highly complex to tackle due to its non-convex nature and the difficulty in treating any generic form of the pdf $f(\vv{w})$. In the following, we propose a simple and efficient solution based on Monte Carlo sampling and SDR.


\subsection{Proposed canonical solution (\name{})}\label{subsec:fixed_drone_sol}

Let $\overline{a}_{BS}(\vv{v}) = \vv{a}_{BS}^{\herm}(\vv{q})\vv{v}$ such that we can simplify the numerator of the objective function of Problem~\ref{eq:Prob} as follows
\begin{align}
    &\Exp[|\vv{a}_T^{\herm}(\vv{q},\vv{w},\vv{r})\vv{\Theta}\vv{a}_R(\vv{q},\vv{r})\vv{a}_{BS}^{\herm}(\vv{q})\vv{v}|^2]\\
    & = |\overline{a}_{BS}(\vv{v})|^2 \Exp[|\vv{a}_T^{\herm}(\vv{q},\vv{w},\vv{r})\vv{\Theta}\vv{a}_R(\vv{q},\vv{r})|^2] \\
    & = |\overline{a}_{BS}(\vv{v})|^2  \Exp\left[\Bigg\rvert\sum_{i=1}^Na_{T,i}^*[\vv{\Theta}]_{ii} a_{R,i}\Bigg\rvert^2\right],\label{eq:espectation_final}
\end{align}
where we have dropped the dependency on $\vv{q}$, $\vv{w}$, and $\vv{r}$ to ease the notation. The term $a_{T,i}^*[\vv{\Theta}]_{ii}a_{R,i}$ in Eq.~\eqref{eq:espectation_final} can be expressed as $[\vv{\Theta}]_{ii}\tilde{a}_{i}$, with $\tilde{a}_{i} = a_{T,i}^*a_{R,i}$. Note that $|\tilde{a}_{i}| = 1$ and $\angle \tilde{a}_{i} = \tilde{\phi}_{i} = \phi^T_{i} - \phi^R_{i}$.
Let $\vv{\theta} = \mathrm{diag}(\vv{\Theta})$ such that we can rewrite Eq.~\eqref{eq:espectation_final} as
\begin{align}
    &|\overline{a}_{BS}(\vv{v})|^2  \Exp\left[\rvert \vv{\theta}^{\tran}\tilde{\vv{a}}(\vv{q},\vv{w},\vv{r})\rvert^2\right]\\
    & = |\overline{a}_{BS}(\vv{v})|^2  \Exp\left[\vv{\theta}^{\tran}\tilde{\vv{a}}(\vv{q},\vv{w},\vv{r})\tilde{\vv{a}}^{\herm}(\vv{q},\vv{w},\vv{r})\vv{\theta}^*\right]\\
    & = |\overline{a}_{BS}(\vv{v})|^2  \vv{\theta}^{\tran}\Exp\left[\tilde{\vv{a}}(\vv{q},\vv{w},\vv{r})\tilde{\vv{a}}^{\herm}(\vv{q},\vv{w},\vv{r})\right]\vv{\theta}^*\\
    & = |\overline{a}_{BS}(\vv{v})|^2  \vv{\theta}^{\tran}\widetilde{\vv{A}}(\vv{q},\vv{w})\vv{\theta}^*,
\end{align}
where the elements of the matrix $\widetilde{\vv{A}}(\vv{q},\vv{w})$ are defined as $[\widetilde{\vv{A}}(\vv{q},\vv{w})]_{i\ell} = \Exp\left[\tilde{a}_i\tilde{a}^*_\ell\right] = \Exp\left[e^{j(\tilde{\phi}_{i}-\tilde{\phi}_{\ell})}\right]$. We assume that the matrix $\widetilde{\vv{A}}(\vv{q},\vv{w})$ is approximated (we refer the reader to Appendix~\ref{ap:A1} for more details) or estimated at the BS-side via a separate control channel.

The phases $\{\tilde{\phi}_{i}-\tilde{\phi}_{\ell}\}_{i,\ell = 1}^N$ corresponding to the element in position $(i,\ell)$ of matrix $\widetilde{\vv{A}}(\vv{q},\vv{w})$ can be expressed as a function of $\vv{r}$ as the following
\begin{align}
    &\tilde{\phi}_{i} - \tilde{\phi}_{\ell} \nonumber\\
    & = \frac{2\pi}{\lambda}\big[\left(\eta_1 \cos{\psi_z} + \eta_2 \sin{\psi_z} \right)\cos{\psi_y}+ \eta_{3} \sin{\psi_y}\big](x_i-x_{\ell}) \nonumber\\
    & + \frac{2\pi}{\lambda}\big[\left(\eta_2 \cos{\psi_z} - \eta_1 \sin{\psi_z} \right)\cos{\psi_x} + \eta_{3} \sin{\psi_x} \big](y_i-y_{\ell})\label{eq:phi_tilde_i_j},
\end{align}
with
\begin{align}
    \eta_1 &= \cos{\theta_T}\cos{\phi_T} - \cos{\theta_R}\cos{\phi_R},\label{eq:eta_1}\\
    \eta_2 &= \cos{\theta_T}\sin{\phi_T} - \cos{\theta_R}\sin{\phi_R}\label{eq:eta_2},
    \end{align}
and
\begin{align}
    \eta_3 &= \sin{\theta_T} - \sin{\theta_R}.\label{eq:eta_3}
\end{align}
We then reformulate Problem~\ref{eq:Prob} as the following

\begin{align}
\begin{array}{cl}
    \displaystyle \max_{\vv{\theta},\vv{v}} \  \min_{\substack{\vv{w}\in \mathcal{A} \\ \vv{w}\sim f_{\vv{w}}(\vv{w})}} & \displaystyle \frac{|\vv{a}_{BS}^{\herm}(\vv{q})\vv{v}|^2 \vv{\theta}^{\tran}\widetilde{\vv{A}}(\vv{q},\vv{w})\vv{\theta}^*}{\|\vv{q}-\vv{w}\|^2} \\
     \mathrm{s.t.} & \quad \|\vv{v}\|^2\leq P;\\
        & \quad |\theta_i|^2\leq 1, \ \forall i.
\end{array}
\end{align}
Note that in the above the optimal value of the precoder $\vv{v}$ adopted at the BS does not depend on $\vv{w}$ and, when the vector of the RIS parameters $\vv{\theta}$ is kept constant, can be optimized by aligning it to the direction of departure of the link between the BS and the RIS. Hence, by noting that $\|\vv{a}_{BS}(\vv{q})\|^2 = M$ and by using maximum ratio transmission (MRT) we set
\begin{align}
    \vv{v} = \sqrt{\frac{P}{M}}\vv{a}_{BS}(\vv{q}).\label{eq:MRT}
\end{align}
We are thus left with the following optimization problem on RIS parameters
\begin{problem}[P\_{FD}]\label{eq:P_optim_theta}
\begin{align}
        \displaystyle \max_{\vv{\theta}} \  \min_{\substack{\vv{w}\in \mathcal{A} \\ \vv{w}\sim f_{\vv{w}}(\vv{w})}} & \displaystyle \frac{ \vv{\theta}^{\tran}\widetilde{\vv{A}}(\vv{q},\vv{w})\,\vv{\theta}^*}{\|\vv{q}-\vv{w}\|^2} \\
        \mathrm{s.t.} & |\theta_i|^2\leq 1 \ \forall i.
\end{align}
\end{problem}
Problem~\ref{eq:P_optim_theta} is still non-convex due to the maximization of a quadratic function in $\vv{\theta}$. An efficient solution can be found by employing SDR as detailed in the following. Let $\overline{\vv{\Theta}} \triangleq \vv{\theta}\vv{\theta}^{\herm}$, which allows us to reformulate Problem~\ref{eq:P_optim_theta} as the following
\begin{align}
    \begin{array}{cl}\label{eq:P_optim_Thetab_1}
        \displaystyle \max_{\vv{\overline{\Theta}}\succeq\vv{0}} \  \min_{\substack{\vv{w}\in \mathcal{A} \\ \vv{w}\sim f_{\vv{w}}(\vv{w})}} & \displaystyle \frac{ \tr(\widetilde{\vv{A}}^*(\vv{q},\vv{w})\,\overline{\vv{\Theta}})}{\|\vv{q}-\vv{w}\|^2} \\
        \mathrm{s.t.} & \mathrm{diag}(\overline{\vv{\Theta}}) \leq \vv{1};\\
        & \mathrm{rank}(\overline{\vv{\Theta}}) = 1.
\end{array}
\end{align}
In order to cope with any given general probability density function of the distribution of the receivers $f_{\vv{w}}(\vv{w})$, we propose to apply a Monte Carlo sampling approach whereby we drop a number of sample points $N_w$ within the target area $\mathcal{A}$ and according to a-priori statistics $f_{\vv{w}}(\vv{w})$. If $N_w$ is sufficiently high, we obtain a correct sampling of the pdf of the receivers in the target area. Note that $N_w$ is expressed as a trade-off between complexity and accuracy of the proposed approach. Let $\overline{\mathcal{A}}$ be the set of such points such that the above problem can be solved by standard semidefinite programming (SDP) (e.g., CVX) in its relaxed convex form, i.e., by ignoring the non-convex rank constraint. The solution obtained by SDP can be projected onto a rank-one space by Gaussian randomization or eigenvalue decomposition~\cite{Luo10}. We thus find an approximate solution to Problem~\ref{eq:P_optim_theta} by solving the following
\begin{problem}[P\_FD\_SDR]\label{eq:P_optim_Thetab}
\begin{align}
        \displaystyle \max_{\vv{\overline{\Theta}}\succeq\vv{0}} \  \min_{\vv{w}\in \mathcal{\overline{A}}} & \displaystyle \frac{ \tr(\widetilde{\vv{A}}^*(\vv{q},\vv{w})\,\overline{\vv{\Theta}})}{\|\vv{q}-\vv{w}\|^2} \\
        \mathrm{s.t.} & \quad \mathrm{diag}(\overline{\vv{\Theta}}) \leq \vv{1}.
\end{align}
\end{problem}
The proposed procedure is formally described in Algorithm \ref{alg:FD_Alg}. \change{The complexity of \name{} is essentially dictated by step~\ref{step:sdp}, i.e., the solution of Problem~\ref{eq:P_optim_Thetab}. The latter is typically implemented via standard SDP such as the bisection method, whose convergence is guaranteed thanks to the convex nature of Problem~\ref{eq:P_optim_Thetab}. Such method typically requires $10$ to $12$ iterations each one having a complexity of $\mathcal{O}(\sqrt{N}(N^6+N^3))$. Moreover, the complexity of the Gaussian randomization in step~\ref{step:gauss_rand} is negligible compared to the previous step \cite{Kar08}.}

\begin{algorithm}[t!]
  \caption{RiFe}\label{alg:FD_Alg}
  \begin{algorithmic}[1]
     \State Initialize $N_w$
     \State Form set $\overline{\mathcal{A}}$ by dropping $N_w$ points $\{\vv{w}_i\}_{i=1}^{N_w}$ according to $f_{\vv{w}}(\vv{w})$
     \State Solve Problem \ref{eq:P_optim_Thetab} and obtain $\overline{\vv{\Theta}}$ \label{step:sdp}
     \State Extract $\vv{\theta}$ from $\overline{\vv{\Theta}}$ via Gaussian randomization\label{step:gauss_rand}
  \end{algorithmic}
\end{algorithm}

\section{Practical considerations}\label{sec:practical_cons}

We further propose a more practical solution aiming at simplifying the optimization procedure described in Algorithm~\ref{alg:FD_Alg}. Specifically, Problem~\ref{eq:P_optim_Thetab} requires to be solved every time the statistics of $\vv{r}$ are no longer valid which, given the need to use SDP, can be excessively time-consuming when the statistics change rapidly. This motivates us to propose a suboptimal yet simple closed-form solution to Problem~\ref{eq:P_optim_Thetab}. 

\begin{figure}
\subfloat[UAV orientation perturbation $\sigma = 2$°.\label{subfig:rad_pattern_2}]{
     \centering
     \includegraphics[width=0.9\columnwidth]{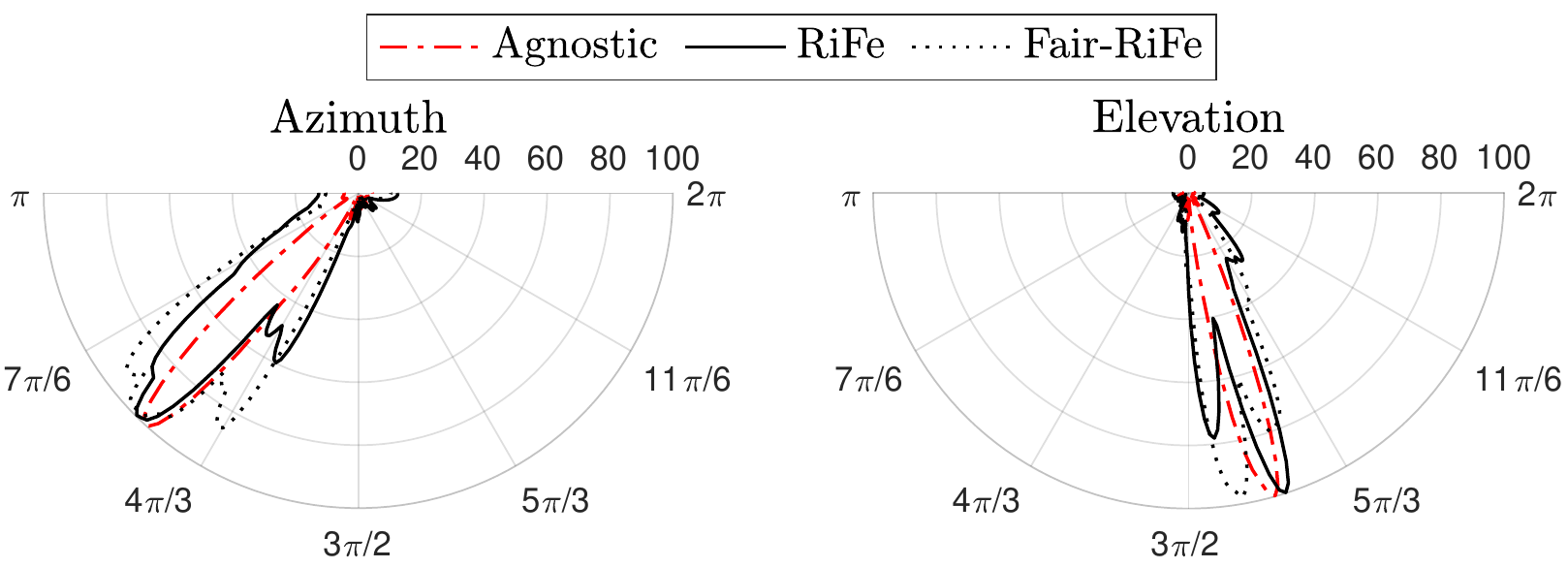}
     }
     \hfill
\subfloat[UAV orientation perturbation $\sigma = 5$°.\label{subfig:rad_pattern_5}]{
     \centering
       \includegraphics[width=0.9\columnwidth]{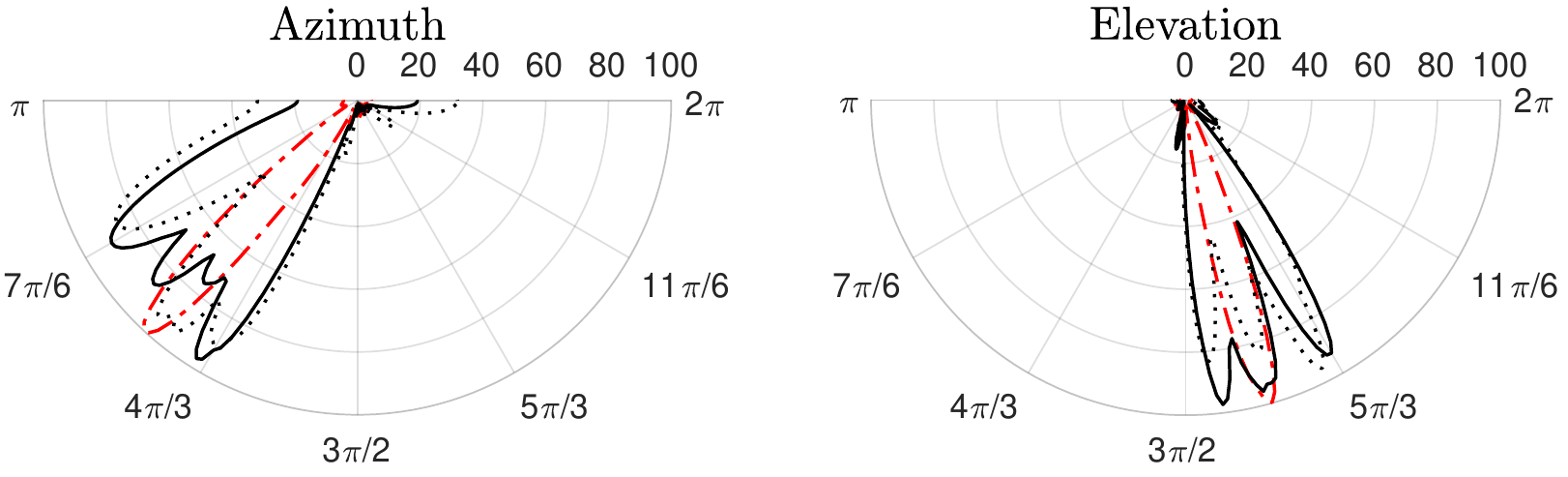}
        }
      \caption{Radiation pattern at the RIS along the azimuth and elevation directions obtained with \name{}, Fair-\name{}, and with the agnostic scheme for different values of the UAV orientation perturbations, where $\sigma = \sigma_{\psi,x}  =\sigma_{\psi,y} = \sigma_{\psi,z}$ and with the default values in Table~\ref{tab:params}.}
           \label{fig:rad_pattern}%
\end{figure}

Given a total number of sampling points $N_w$, the objective is to maximize the worst SNR among the associated ones in set $\overline{\mathcal{A}}$. Hence, we fix the matrix $\overline{\vv{\Theta}}$ to a weighted combination of the $\widetilde{\vv{A}}^*(\vv{q},\vv{w})$ matrices as
\begin{align}\label{eq:Theta_bar_heur}
    \overline{\vv{\Theta}} = \sum_{i=1}^{N_w} \lambda_i \widetilde{\vv{A}}^*(\vv{q},\vv{w}_i).
 \end{align}
The weights $\lambda_i$ need to be chosen such that more power is allocated along the propagation directions corresponding to the point exhibiting the worst SNR w.r.t. the others in set $\overline{\mathcal{A}}$ and such that the resulting matrix $\overline{\vv{\Theta}}$ is a positive semi-definite matrix whose diagonal elements are less than or equal to $1$. Since the received SNR depends largely on the distance from the target point to the UAV location, we set
\begin{align} \label{eq:lambda_i}
    \lambda_i = \frac{1}{\overline{\lambda}} \|\vv{q}-\vv{w}_i\|,
\end{align}
where the constant factor $\overline{\lambda}$ is properly chosen to guarantee the aforementioned constraints on the matrix $\overline{\vv{\Theta}}$. In this regard, note that $\widetilde{\vv{A}}^*(\vv{q},\vv{w}_i)$ is a positive semi-definite matrix and $\lambda_i >0, \,\, \forall i$ thus ensuring that $\overline{\vv{\Theta}}$ is a positive semi-definite matrix as well. Moreover, the value of $\overline{\lambda}$ is found by constraining the diagonal of $\overline{\vv{\Theta}}$ to $\vv{1}$ as
\begin{align}
    \mathrm{diag}(\overline{\vv{\Theta}}) = \sum_{i=1}^{N_w} \lambda_i \vv{1},
\end{align}
which ensures that the RIS is a passive structure overall. Finally, we set
\begin{align}\label{eq:lambda_bar}
    \overline{\lambda} = \sum_{i=1}^{N_w} \|\vv{q}-\vv{w}_i\|.
\end{align}
By substituting Eqs.~\eqref{eq:lambda_i} and \eqref{eq:lambda_bar} into Eq.~\eqref{eq:Theta_bar_heur} we obtain
\begin{align}\label{eq:Theta_bar_heur2}
    \overline{\vv{\Theta}} =
    \frac{1}{\sum_{i=1}^{N_w} \|\vv{q}-\vv{w}_i\|} \sum_{i=1}^{N_w}  \|\vv{q}-\vv{w}_i\| \widetilde{\vv{A}}^*(\vv{q},\vv{w}_i).
 \end{align}
The RIS parameters in the vector $\vv{\theta}$ can be obtained from Eq.~\eqref{eq:Theta_bar_heur2} by Gaussian randomization or eigenvalue decomposition. The proposed procedure is formalized in Algorithm~\ref{alg:FD_Alg_Heur} and is denoted as Fair-\name{}. In Fig.~\ref{fig:rad_pattern} we compare the radiation pattern at the RIS along the azimuth and elevation directions obtained with \name{}, Fair-\name{} and with an agnostic solution that does not take into account the UAV perturbations. As expected, the agnostic solution points towards the center of the target area, regardless of the entity of the drone perturbations. Whereas, both \name{} and Fair-\name{} tend to spread the energy over a wider angular span as the perturbations increase.

\begin{algorithm}[t!]
  \caption{Fair-RiFe}\label{alg:FD_Alg_Heur}
  \begin{algorithmic}[1]
     \State Initialize $N_w$
     \State Form set $\overline{\mathcal{A}}$ by dropping $N_w$ points $\{\vv{w}_i\}_{i=1}^{N_w}$ according to $f_{\vv{w}}(\vv{w})$
     \State Set $\lambda_i = \frac{1}{\overline{\lambda}} \|\vv{q}-\vv{w}_i\|$ $\forall \vv{w}_i \in \overline{\mathcal{A}}$ with $\overline{\lambda} = \sum_{i=1}^{N_w} \|\vv{q}-\vv{w}_i\|$ 
     \State Fix $\overline{\vv{\Theta}} =
     \sum_{i=1}^{N_w}  \lambda_i \widetilde{\vv{A}}^*(\vv{q},\vv{w}_i)$
     \State Extract $\vv{\theta}$ from $\overline{\vv{\Theta}}$ via Gaussian randomization
  \end{algorithmic}
\end{algorithm}

\subsection{Special case: deterministic UAV movements}\label{sec:moving_drone_known}

Let us consider a special case of the above-described scenario, wherein the UAV is moving with given direction and velocity. In this case the statistics of $\vv{r}$ changes rapidly together with drone movements, which requires an updated matrix $\widetilde{A}(\vv{q},\vv{w})$ and the solution of Problem~\ref{eq:Prob}. As a result of the movement of the UAV, the surface of the RIS exhibits an (expected) inclination of $\hat{\psi}_x$, $\hat{\psi}_y$, and $\hat{\psi}_z$ along the $x$, $y$, and $z$ axes, respectively.\footnote{Note that such expected inclination of the surface based on the UAV speed can be readily calculated via flight controller. We refer the reader to~\cite{simma2020measuring} for more details.} In this case, our proposed approach is readily applicable by simply substituting the coefficients in Eqs.~\eqref{eq:eta_1}--\eqref{eq:eta_2} with
\begin{align}
\eta_1 &= \cos{\theta_T} \cos{(\phi_T - \hat{\psi}_z)}-\cos{\theta_R}\cos{(\phi_R - \hat{\psi}_z)},
\end{align}
and
\begin{align}
\eta_2 &= \cos{\theta_T} \sin{(\phi_T - \hat{\psi}_z)}-\cos{\theta_R}\sin{(\phi_R - \hat{\psi}_z)},
\end{align}
respectively. Moreover, in this case the phase term in position $(i,\ell)$ of matrix $\widetilde{\vv{A}}(\vv{q},\vv{w})$ is expressed as
\begin{align}
&\tilde{\phi}_i - \tilde{\phi}_{\ell} =\nonumber\\
&\frac{2\pi}{\lambda}\big[(\eta_2\cos{\hat{\psi}_x} \cos{\psi_z}-\eta_1\cos{\hat{\psi}_x} \sin{\psi_z}+\eta_3\sin{\hat{\psi}_x})\cos{\psi_x}\nonumber\\
& - (\eta_2\sin{\hat{\psi}_x} \cos{\psi_z}-\eta_1\sin{\hat{\psi}_x} \sin{\psi_z}-\eta_3\cos{\hat{\psi}_x})\sin{\psi_x}\big]\nonumber\\
& \times (y_i-y_{\ell})-\frac{2\pi}{\lambda}\big[(\eta_1\cos{\hat{\psi}_y} \cos{\psi_z}+\eta_2\cos{\hat{\psi}_y} \sin{\psi_z}\nonumber\\
& +\eta_3\sin{\hat{\psi}_y})\cos{\psi_y}- (\eta_1\sin{\hat{\psi}_y} \cos{\psi_z}+\eta_2\sin{\hat{\psi}_y} \sin{\psi_z}\nonumber\\
&-\eta_3\cos{\hat{\psi}_y})\sin{\psi_y}\big](x_i-x_{\ell}).
\end{align}
Hence, both Algorithms \ref{alg:FD_Alg} and \ref{alg:FD_Alg_Heur} can be alternatively applied without any further modification. 

\section{Performance evaluation}\label{sec:Perf_eval}

We present numerical results to assess the performance of our proposals by means of ad-hoc MATLAB\textregistered\, simulations. We place one BS with $M=16$ antenna elements at coordinates $(0,0,10)$~m, while the RIS is installed on a drone that is located at coordinates $(25,25,h_d)$~m, where the altitude of the drone is set to $h_d=50$~m, unless otherwise stated. The RIS is equipped with $N_x \times N_y$ antenna elements, with $N_x=N_y=10$, unless otherwise stated. Hence, the total number of RIS elements is $N=N_xN_y = 100$ with an inter-distance of $d=\frac{\lambda}{2}$.\footnote{Note that in our simulations we consider that an inter-distance of $\lambda/2$ allows to neglect the mutual coupling effect between RIS elements (c.f.~\cite{Qia20}).} The target area wherein first responder teams and/or victims are located is centered at coordinates $(50,20,0)$~m. The BS transmit power is set to $P=24$~dBm and the average noise power to $\sigma_n^2 =-80$~dBm. We assume that the variances of the perturbations of the drone orientation along the $x$, $y$, and $z$ axes are equal to $\sigma_{\psi_x}=\sigma_{\psi_y}=\sigma_{\psi_z}=\sigma$, with $\sigma$ ranging from $0^\circ$ to $20^\circ$. Lastly, the working frequency is set to $f=30$~GHz and the channel gain at a reference distance of $1$~m is assumed to be equal to $\beta_0=1$. All simulation settings are listed in Table~\ref{tab:params}, unless otherwise stated. The numerical results for each simulation instance are averaged over $5000$ realizations of the drone perturbations, i.e., of the vector $\vv{r}$, and the number of sample points necessary for Monte Carlo sampling of the distribution of the receivers is set to $N_w = 200$. Lastly, we average the obtained results over $10$ instances of random locations of the receivers in the target area. Numerical simulations are carried out on an Intel i7-8665U CPU @ $1.90$~GHz and $32$~GB RAM $8$-core machine.

\begin{table}[h!]
\caption{Default simulation settings}
\label{tab:params}
\centering
\resizebox{.85\linewidth}{!}{%
\begin{tabular}{c|c|c}
\textbf{Parameter} & \textbf{Description} & \textbf{Default Value}\\  
\hline
\rowcolor[HTML]{EFEFEF}
$\vv{q}$ & Drone coordinates & $(25,25,h_d)$ \\
$h_d$ & Drone altitude & $50$ m \\
\rowcolor[HTML]{EFEFEF}
$\vv{b}$ & BS coordinates & $(0,0,10)$ \\
$\vv{c}_{A}$ & Center of target area & $(50,20,0)$ \\
\rowcolor[HTML]{EFEFEF}
$N$  &  Number of RIS elements & $100$  \\
$M$  &  Number of BS antenna elements & $16$  \\
\rowcolor[HTML]{EFEFEF}
$P$ & BS transmission power & $24$ dBm \\
$\sigma^2_n$ & Noise power & $-80$ dBm \\
\rowcolor[HTML]{EFEFEF}
$f$ & Working frequency & $30$ GHz \\
$d$ & RIS element spacing & $\frac{\lambda}{2}$ \\
\rowcolor[HTML]{EFEFEF}
$\sigma$ & Drone orientation perturbation & $20^{\circ}$ \\
$\beta_0$ & Channel gain at $1$ m & \change{$-30$ dB}\\
\end{tabular}%
}
\end{table}

\subsection{Robustness to perturbations}

We first consider the case where the location of the first responder team in the area is perfectly known and the drone is affected by orientation perturbation: we highlight the robustness of our method against the uncertainty on the drone orientation. We compare \name{} against an agnostic procedure that optimizes the RIS parameters by neglecting such perturbations and assuming a perfectly-stable flight, denoted as \emph{agnostic}. We consider $10$ users uniformly dropped in a circular target area of radius {$R=\{20, 35\}$~m}.

\begin{figure}
     \centering
      \subfloat[Drone altitude $h_d=50$ m.\label{fig:min_snr_osc_h1}]{%
       \includegraphics[width=0.485\columnwidth]{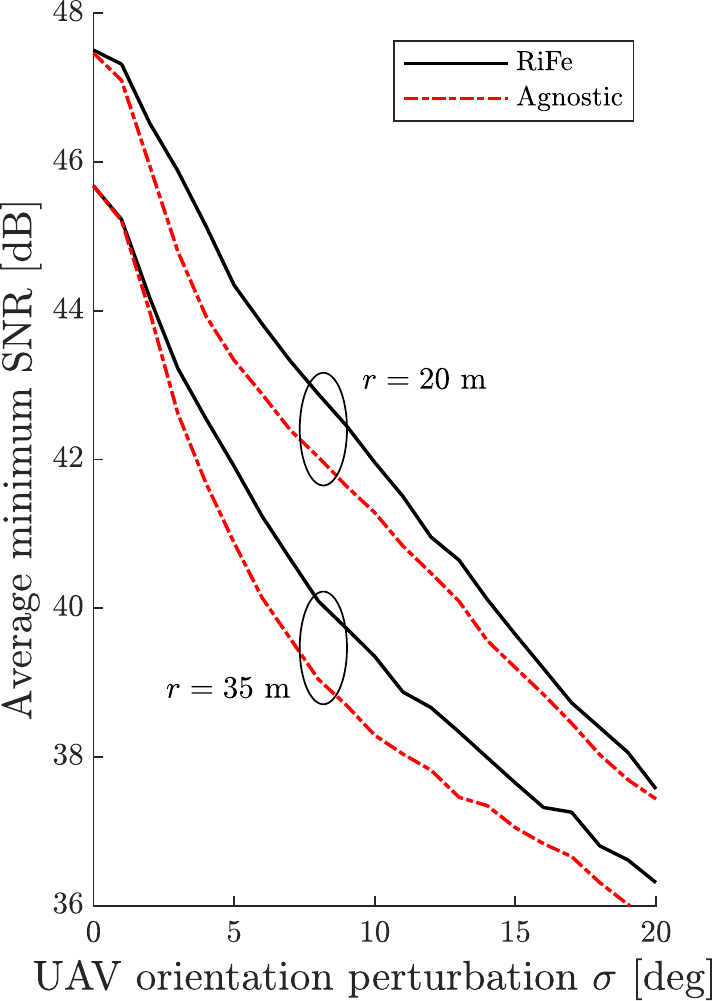}
   }
   \hfill
          \subfloat[Drone altitude $h_d=70$ m.\label{fig:min_snr_osc_h2}]{%
       \includegraphics[width=0.485\columnwidth]{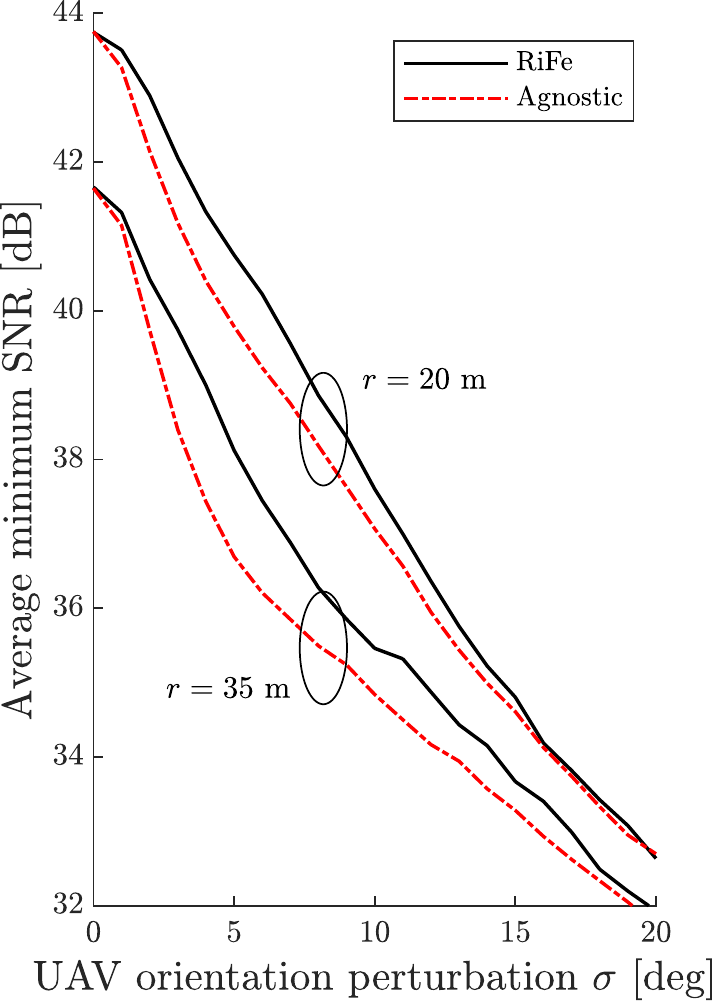}
   } 
        \caption{\change{Robustness to drone perturbations: average minimum SNR obtained with \name{} and with the agnostic solution for different values of the radius $r$ of the target area and the drone altitude $h_d$.}}
        \label{fig:min_snr_osc}
\end{figure}


\begin{figure}
     \centering
      \subfloat[Drone altitude $h_d=50$ m.\label{fig:cdf_osc_h1}]{%
       \includegraphics[width=0.463\columnwidth]{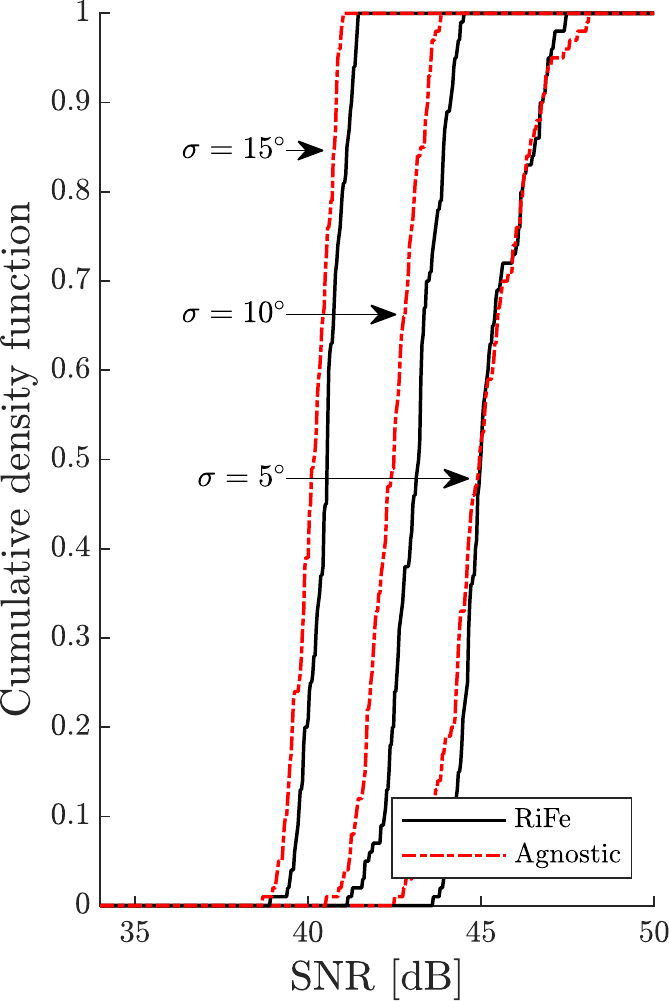}
   }
   \hfill
          \subfloat[Drone altitude $h_d=70$ m.\label{fig:cdf_osc_h2}]{%
       \includegraphics[width=0.463\columnwidth]{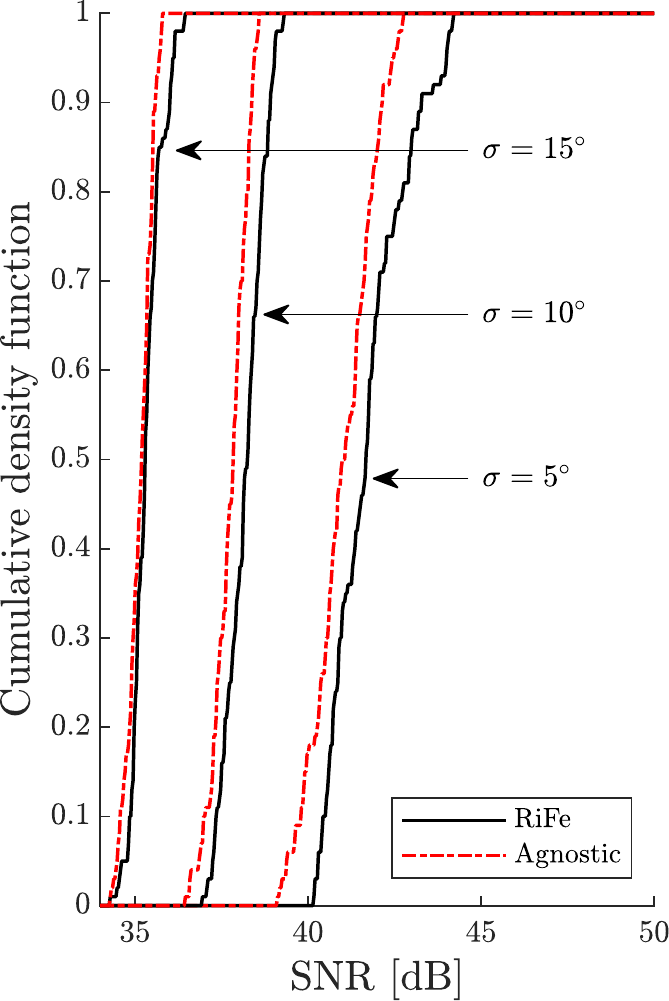}
   } 
        \caption{\change{Trade-off between performance improvement and perturbation magnitude: CDF of the received SNR within a target area of radius $r=35$ m obtained with \name{} and the agnostic solution for different values of the UAV orientation perturbation $\sigma$ and the drone altitude $h_d$.}}
        \label{fig:cdf_snr_osc}
\end{figure}


Fig.~\ref{fig:min_snr_osc} compares the performances of \name{} against the agnostic scheme in terms of the minimum SNR received by the first responder team in the target area. It is worth noting that drone oscillations results in a reduction of the received SNR, which is mainly caused by the misalignment of the reflected beams (on the RIS) with respect to the users to be covered in the target area. Notably, \name{} mitigates such misalignment effects by optimizing the RIS parameters as a function of the statistics of such oscillations. Indeed, our proposed approach tends to generate wider beams compared to the agnostic solution, i.e., it better spreads the radiated power over the target area thus compensating an undesired misalignment.

Interestingly, \name{} performance gains are less evident as the dimension of the target area increases. When points to be covered are well separated, reflected beams on the RIS focus on each target point. Vice versa, if the receivers are in close proximity then \name{} tends to cover all the target points with a single wide beam. This results in a similar RIS configuration compared to the agnostic solution, thus reducing the overall performance gain in terms of received SNR.

Additionally, we observe a trade-off between the effectiveness of the oscillation compensation mechanism of \name{} and the magnitude of such oscillations. Indeed, the two considered methods produce similar gains in the two extreme cases when the magnitude of the perturbations is low because of a relatively low drone altitude or small values of $\sigma_{\psi}$, and when the size of perturbations is large because of high altitude of the drone or a large value of $\sigma_{\psi}$. The rationale behind this behavior is the following: when perturbations are limited, their corresponding effect is negligible, thus the improvement of the proposed compensation mechanism is negligible. Conversely, when the altitude is high, small oscillations may result in large deviations of the corresponding beams with respect to the desired pointing direction. In this case, \name{} will generate wide beams to spread the irradiated power over a larger angular span thereby resulting in an inevitable reduction of the received signal power. Such a behavior is highlighted in Fig.~\ref{fig:cdf_snr_osc}, which shows the cumulative density functions (CDF) of the received SNR.

\begin{figure}
\centering
\includegraphics[width=0.8\columnwidth]{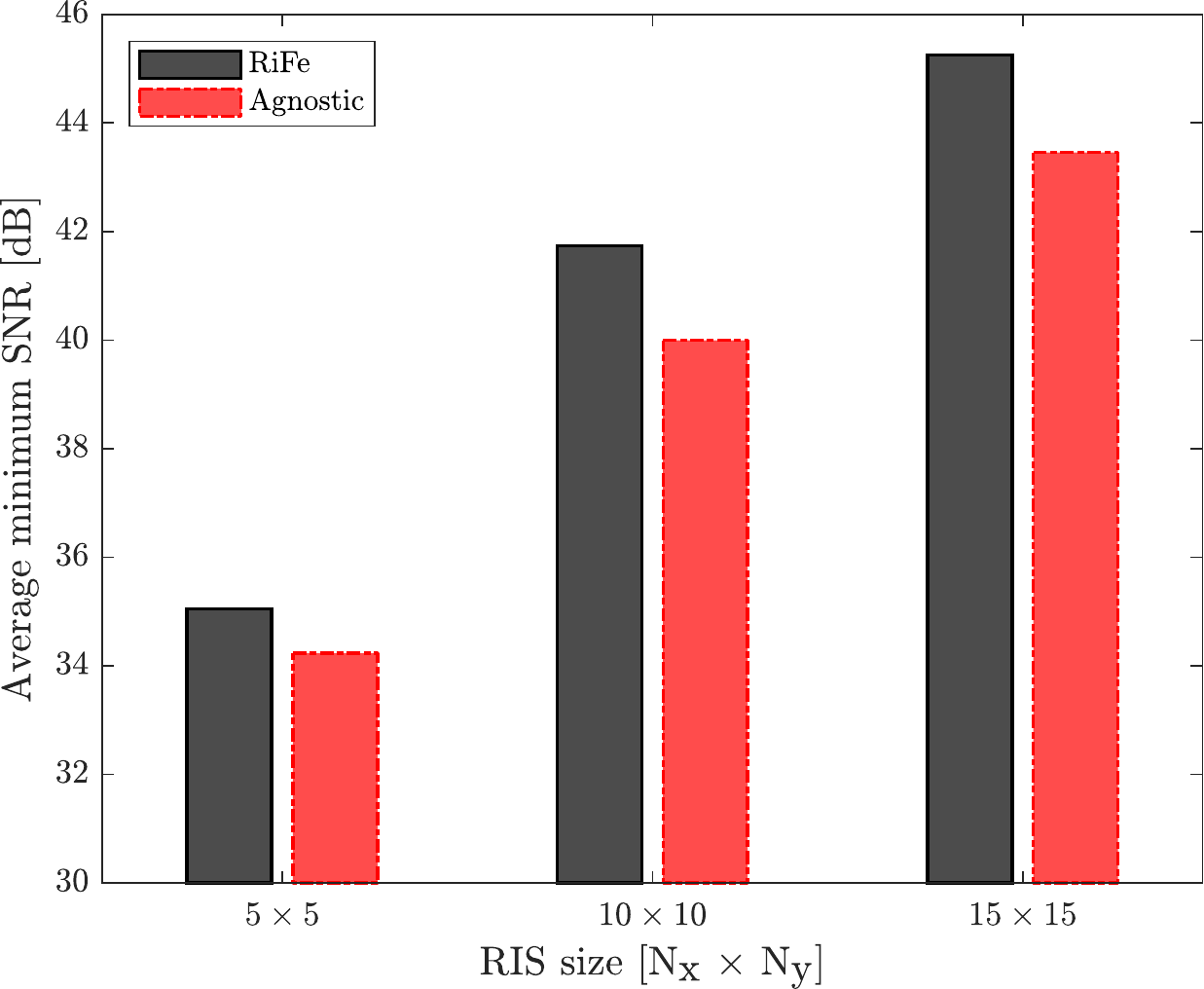}
	\caption{\change{Robustness to drone perturbations: average minimum SNR obtained with \name{} and with the agnostic solution for different setting of the number of RIS elements $N$ with the UAV altitude set to $h_d = 50$~m, a target area with radius $r=35$~m, and UAV perturbation $\sigma=5^{\circ}$.}}
	\label{fig:avg_min_snr_ris_el}
\end{figure}

We perform additional tests to take into account the effect of the RIS size by varying the number of elements $N$. Fig.~\ref{fig:avg_min_snr_ris_el} compares the performances of \name{} against the agnostic scheme in terms of minimum SNR. It is worth noticing the overall increasing trend of the achieved minimum SNR according to the number of elements. This due to the ability to focus---which is proportional to its dimension---that reflects on the directionality of the reflected beams. However, the more the energy is focused towards the target points, the higher the effect of misalignment due to the drone orientation perturbation. This results, in turn, in the gap between the performances of \name{} and agnostic optimization schemes, which increases according to the available number of elements, thus highlighting the importance of a robust RIS configuration to combat drone perturbations.

\change{Lastly, as a means to further assess the performance of the proposed \name{} algorithm and the agnostic scheme in terms of worst-case SNR within the target region, we consider the multicast rate, i.e., the rate given by the UE with the worst effective channel condition. This is given by the following
\begin{align}
    R_{\rmmult} \triangleq \log_2\left(1+\min_{\vv{w}\in \mathcal{A}}\mathrm{SNR}(\vv{q,w,r,\Theta,v})\right)\label{eq:multicast_rate},
\end{align}
where the instantaneous SNR for a given drone position $\vv{q}$ and perturbation $\vv{r}$ is given by Eq.~\eqref{eq:snr_2}, $\vv{\Theta}$ is given by the proposed \name{} algorithm, and $\vv{v}$ is set as per Eq.~\eqref{eq:MRT}. The multicast rate in Eq.~\eqref{eq:multicast_rate} is an important key performance indicator in emergency scenarios, considering the need to propagate useful information to the overall first responder team in the shortest time.}
We report the obtained comparison results in Table~\ref{tab:sumrate} to emphasize the benefits of robust optimization strategies to improve the overall system performance and effectiveness.

\begin{table}[h!]
\caption{Multicast rate performance in bps/Hz considering different target area radii and UAV orientation perturbations. Bold and plain values in brackets indicate the performance of RiFe and the agnostic optimization scheme, respectively.}
\label{tab:sumrate}
\centering
\resizebox{.8\linewidth}{!}{%
\begin{tabular}{|c|c|c|c|}
\hline
{Target Area} & \multicolumn{3}{c|}{Perturbation -  {\textbf{\name{}} (Agnostic) }}  \\
 \hline
 & \textit{Low} & \textit{Medium} & \textit{High}\\  
\hline
\rowcolor[HTML]{EFEFEF}
 Radius = $20$ m & \change{$\mathbf{15}$ $(14)$}    & \change{$\mathbf{14}$ $(13)$}     & \change{$\mathbf{14}$ $(13)$} \\
 Radius = $35$ m & \change{$\mathbf{14}$ $(13)$}    & \change{$\mathbf{13}$ $(12)$}     & \change{$\mathbf{13}$ $(12)$}\\
\hline
\end{tabular}%
}

\end{table}

\subsection{Analysis of location-unaware solutions}
\label{sec:Perf_eval_location-unaware}

When the BS aims at serving a number of victims whose position is unknown, our objective is to provide a minimum SNR within the target area to ensure an adequate signal coverage. We assume that the distribution of the victims in the target area is known and follows a symmetric and independent bivariate normal distribution centered in a given point $C$ and with associated standard deviations along the $x$ and $y$ axes equal to $\sigma_{w,x} = \sigma_{w,y} = \sigma_{w}$. The value of $\sigma_{w}$ is set to $\{2.5,10\}$~m, which represent two possible situations, namely victims located in a predefined collection point, and a more sparse distribution around the disaster location. Note that the values of $\sigma_{w}$ correspond to an area of $80$~m$^2$ and $1200$~m$^2$, respectively wherein the $95\%$ of the victims are located. 
Additionally, we compare our solution against 3D beam flattening, which is described in~\cite{Lu20}. 


\begin{figure}
     \centering
      \subfloat[User distribution $\sigma_w=2.5$ m. \label{fig:avg_min_snr_w1}]{%
       \includegraphics[width=0.485\columnwidth]{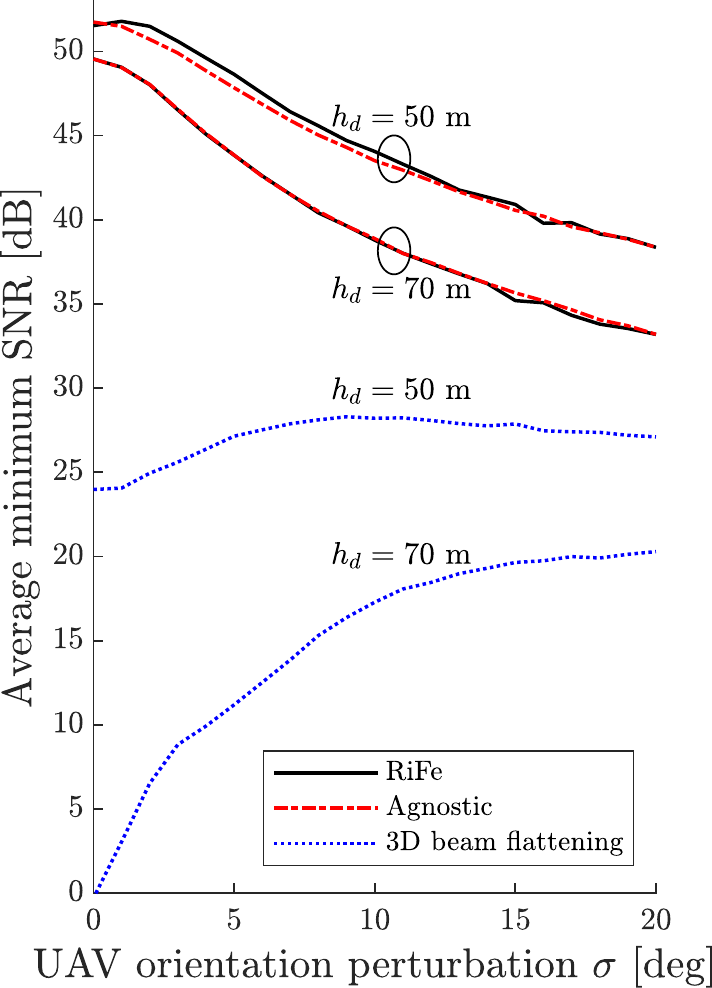}
   }
   \hfill
          \subfloat[User distribution $\sigma_w=10$ m.\label{fig:avg_min_snr_w2}]{%
       \includegraphics[width=0.485\columnwidth]{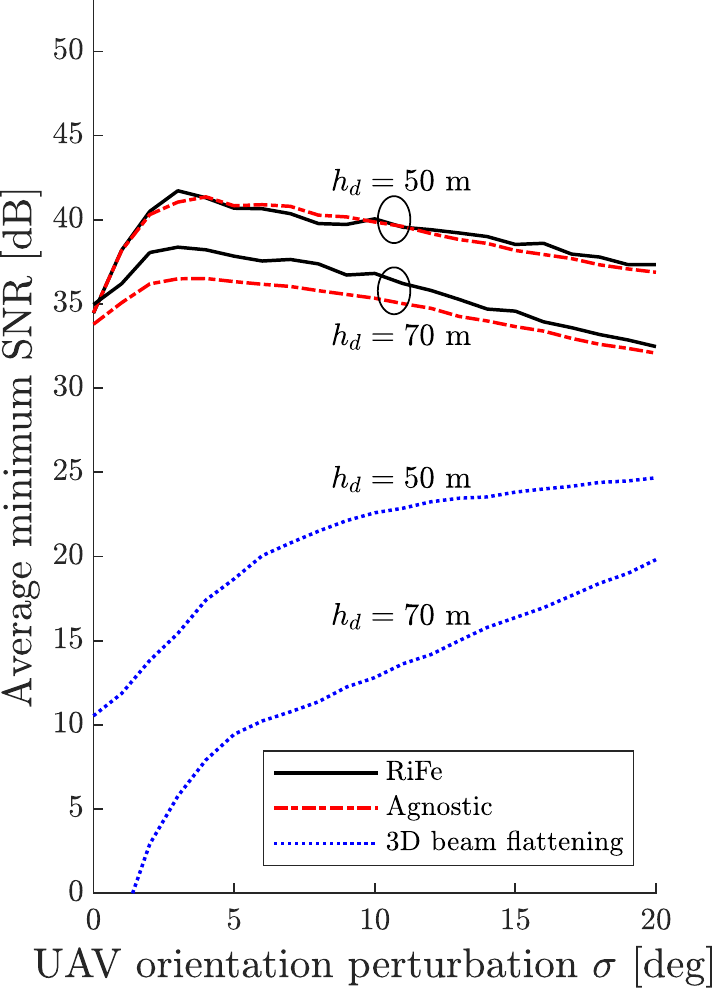}
   } 
        \caption{Average minimum SNR obtained with \name{}, the agnostic solution and with 3D beam flattening \cite{Lu20} over $95\%$ of the users served for different values of the drone altitude $h_d$ and the user spread $\sigma_w$.}
        \label{fig:avg_min_snr}
\end{figure}


\begin{figure}
     \centering
      \subfloat[$\sigma_w=2.5$ m \label{fig:cdf_osc_loc_unaware_h1}]{%
       \includegraphics[width=0.463\columnwidth]{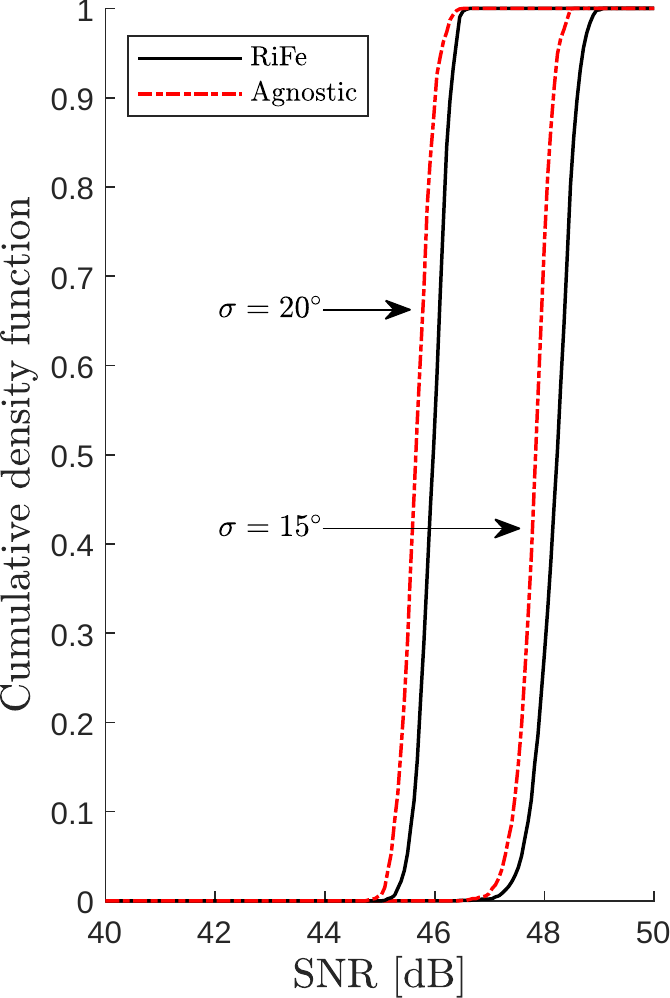}
   }
   \hfill
          \subfloat[$\sigma_w=10$ m\label{fig:cdf_osc_loc_unaware_h2}]{%
       \includegraphics[width=0.463\columnwidth]{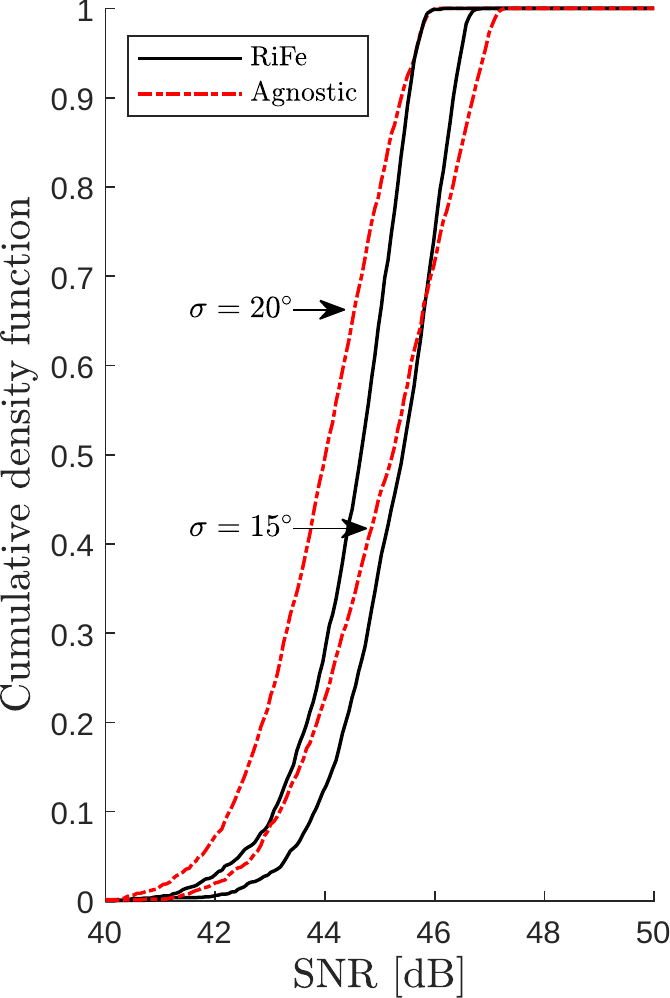}
   } 
        \caption{\change{Trade-off between performance improvement and perturbation magnitude: CDF of the received SNR obtained with \name{} and the agnostic solution with the drone at altitude $h_d=50$~m and for different values of victim spread $\sigma_{w}$.}}
        \label{fig:cdf_snr_osc_loc_unaware}
\end{figure}

Fig.~\ref{fig:avg_min_snr} shows the achieved minimum SNR over the victims dropped in the area with different values of the drone perturbation, altitude of the drone and spread of the victims in the target area.
After comparing the curves with low and high spread of victims in the area, we note that the minimum received SNR has a decreasing trend as the drone perturbation increases. On the other hand, if the victims are more spread, the performance exhibits an increasing trend in the low perturbation regime. This behavior is due to the limited number of samples adopted for the Monte Carlo sampling of the distribution of the victims. Indeed, in the case of low spread of the location of the victims, the proposed method accurately represents the underlying distribution, which in turn results in an effective coverage of the area. Conversely, in the case of large spread of the locations of the victims, the limited number of samples results in a more sparse sampling of the distribution. As a result, the RIS configuration tends to focus the reflected energy towards the available samples instead of following the actual input distribution.

We also note when the victims are widely spread, the beam patterns generated by \name{} result in a considerable SNR improvement since they are sufficiently spread to effectively cover the target area. On the other hand, when the victims are more concentrated, all considered algorithms produce beams that properly cover the target area. However, we can observe an overall increase of the SNR obtained with \name{}. To dig deeper into this behavior, we report in Fig.~\ref{fig:cdf_snr_osc_loc_unaware} the CDF experienced by the victims in the target area considering different considered scenarios. As expected, the CDFs obtained with \name{} have a more pronounced slope with respect to the ones obtained with the agnostic 
solution because of larger beam patterns.


\color{black}

\subsection{Practical evaluation}
In the previous numerical evaluation, we demonstrated the effectiveness of \name{} to obtain a robust RIS configuration, which is able to mitigate the effect of the inherent instability of the drone in flight.
We conclude our analysis by considering the practical implications of the proposed method. To this end, we compare the performance of \name{} algorithm with the Fair~\name{} solution described in Section~\ref{sec:practical_cons}, in terms of target area coverage quality as well as computational cost of the optimization process. For the stake of completeness, we include an agnostic version of Fair~\name{} in our numerical evaluation, namely \textit{Fair-agnostic}, where Algorithm~\ref{alg:FD_Alg_Heur} is solved neglecting the drone perturbation statistic. We consider the same scenario as per Section~\ref{sec:Perf_eval_location-unaware}.

\begin{figure}
\centering
\vspace{5mm}
\includegraphics[width=0.8\columnwidth]{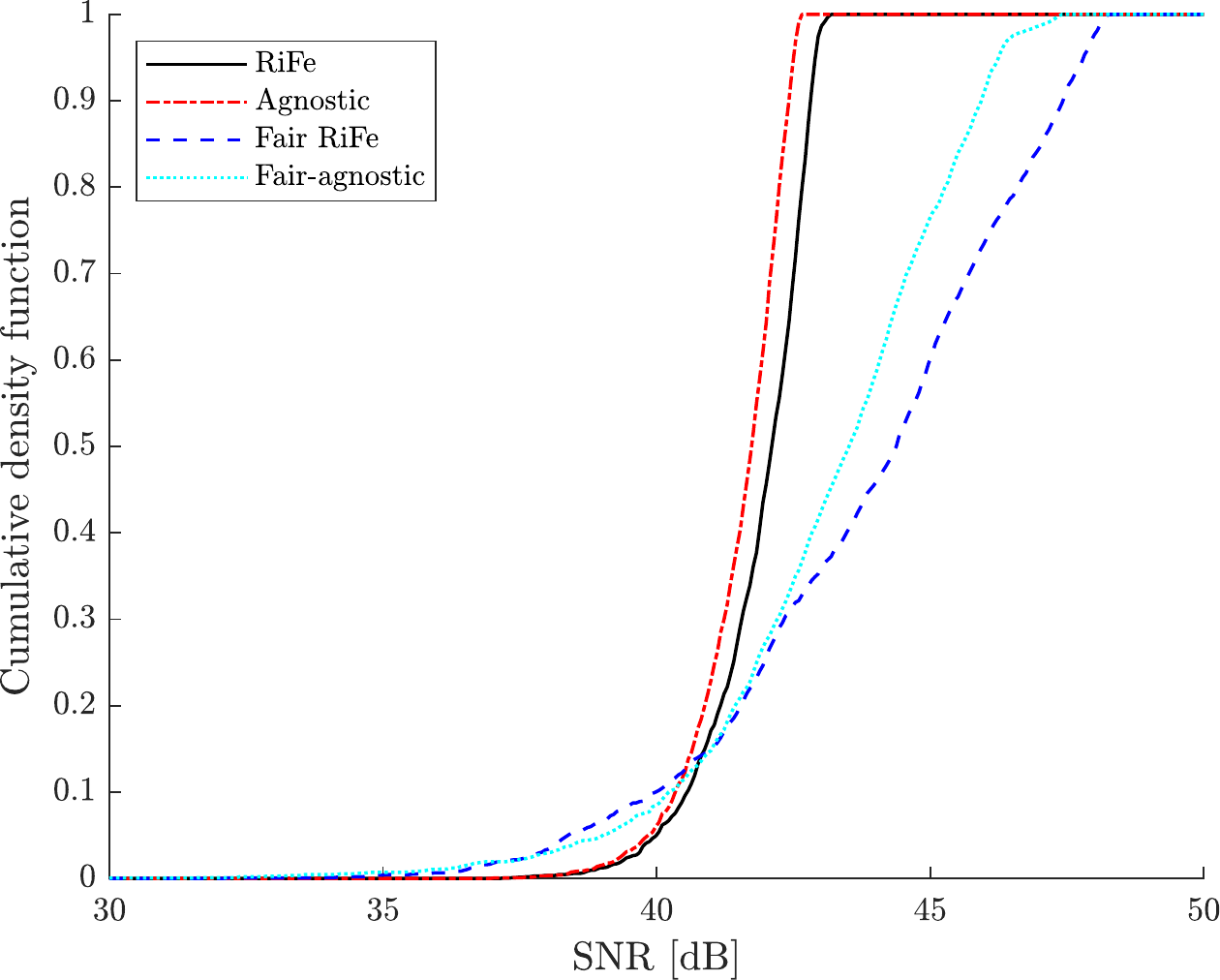}
	\caption{\change{CDF of the received SNR obtained with \name{}, the agnostic solution, Fair~\name{} and the Fair-agnostic scheme with the drone altitude set to $h_d = 50$~m, the victim spread equal to $\sigma_w=2.5$~m, and UAV perturbation $\sigma=5^{\circ}$.}}
	\label{fig:performance_heuristic_vs_solver}
\end{figure}

Fig.~\ref{fig:performance_heuristic_vs_solver} compares the CDF of the victims SNR experienced over a target area considering different proposed solutions. From the results, we can notice that the CDFs obtained with the Fair~\name{} and Fair~agnostic algorithms exhibit a less-marked slope than their optimal counterpart. Thus, they produce a less-homogeneous distribution of the radiated power over the area than their optimal counterpart due to the sub-optimal RIS configuration. Upon comparing the Fair~\name{} and the Fair~agnostic performance, we can notice that, for relatively low SNR values, distributions are very similar to each other, thus leading to similar performances in terms of minimum SNR provided. Vice versa, considering relatively high value of SNR, we can notice that the overall delivered SNR is higher with the Fair~\name{} solution. Therefore, while it delivers a sub-optimal solution, it still provides a performance improvement by considering the drone perturbation statistics in the optimization process.

\begin{figure}
     \centering
      \subfloat[Solution time against number of sample points $N_w$ with $N=10$. \label{fig:solution_time_vs_nw}]{%
      \includegraphics[width=0.485\columnwidth]{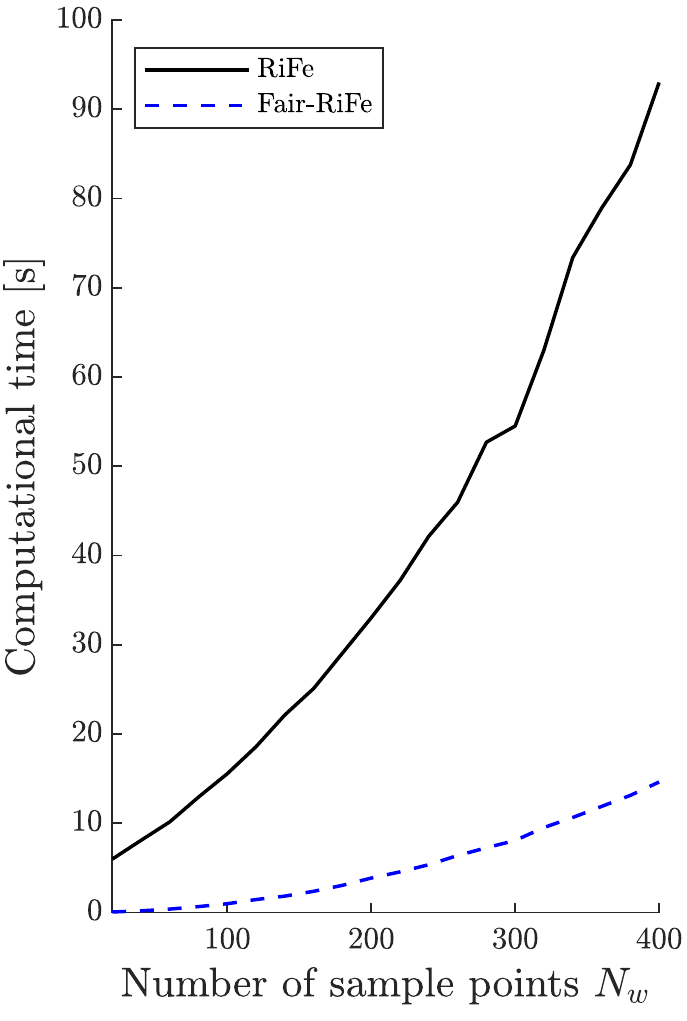}
   }
   \hfill
          \subfloat[Solution time against RIS configuration with $N_w=10$.\label{fig:solution_time_vs_ris_el}]{%
       \vspace{5mm}\includegraphics[width=0.485\columnwidth]{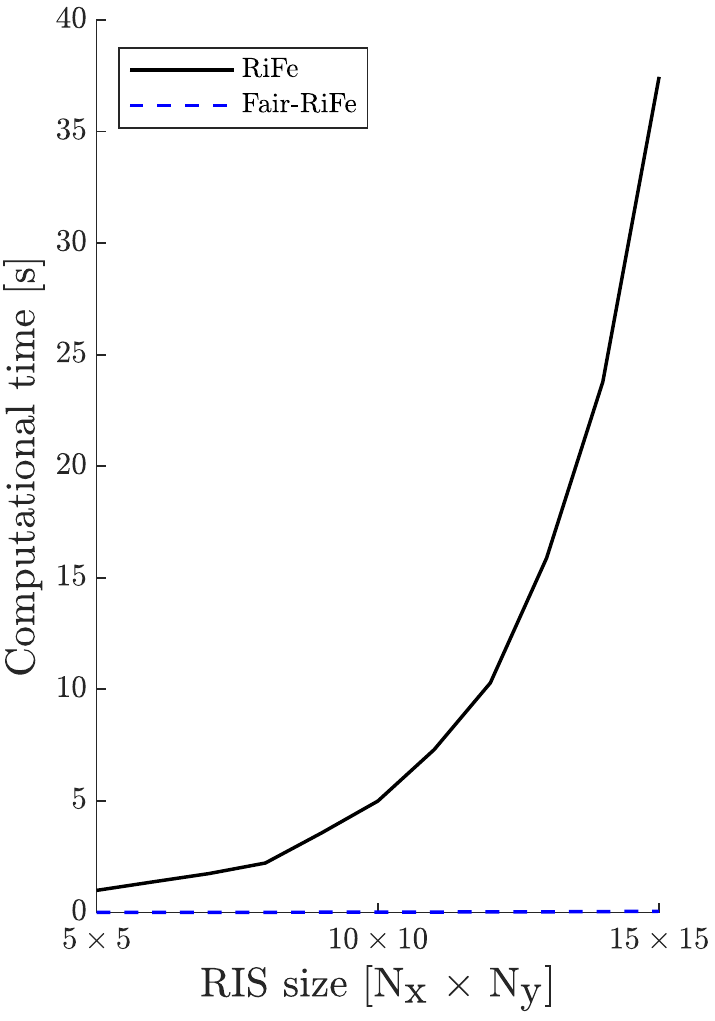}
   } 
        \caption{Comparison of computational time of \name{} and Fair-\name{} algorithms.}
        \label{fig:solution_time}
\end{figure}

Finally, we assess the computational time in Fig.~\ref{fig:solution_time}. In particular, we compare the overall optimization time needed by \name{} and Fair~\name{} solution while varying the number of points within the target area $N_w$ as well as the size of the on-board RIS $N$. 
In Fig.\ref{fig:solution_time_vs_nw} we can notice a dramatic reduction of the optimization time when considering Fair~\name{} against \name{} solution. The reason behind such an improvement relies on the fact that Fair~\name{} does not involve an optimization process to obtain the solution. We can also see that the Fair~\name{} optimization time shows a limited growth w.r.t. the number of points $N_w$ against \name{} solution. Therefore, in the case it is required to sample the target area (i.e., unknown position of the victims) it is suitable for a fine-grained sampling of the victim distribution. Fig.~\ref{fig:solution_time_vs_ris_el} provides the optimization time against the number of RIS elements $N$. Fair~\name{}, which keeps the optimization time almost constant, outperforms \name{}, whose optimization time has an exponential growth w.r.t. the number of RIS elements. Thus, Fair~\name{} allows for a quick configuration of the RIS thereby allowing to exploit the large-sized RIS in real scenario.

\section{Conclusion}\label{sec:concl}

Agile and flexible air-to-ground networks represent the new frontier for reliable communications. In this paper, we envision UAVs equipped with passive devices, namely Reconfigurable Intelligent Surfaces (RISs), able to control the propagation properties of incoming signals to deliver connectivity in emergency scenarios while keeping the energy burden within affordable bounds. 
The main novelty of our \name{} solution is the robustness and reliability of the UAV communication channel by properly combating undesired flight effects, e.g., position perturbations and UAV orientation fluctuations. In addition to the optimization framework, a practical implementation has been designed, Fair-\name{}, which in turn reduces the complexity of the optimal solution. Finally, an exhaustive simulation campaign has been conducted to validate our framework where state-of-the-art solutions are significantly outperformed (e.g., $25$~dB).  

\begin{appendix}

\subsection{Estimating matrix $\widetilde{A}(\vv{q},\vv{w})$}\label{ap:A1}

In the following we describe a possible way to estimate matrix $\widetilde{A}(\vv{q},\vv{w})$ by assuming that the oscillations of the UAV are small, i.e., that $\psi_x,\,\psi_y,\,\psi_z \approx 0$ and using a Taylor expansion such that $\cos(\psi)\approx 1-\psi^2/2$ and $\sin(\psi)\approx \psi$. Let $\Delta x_{i\ell}=x_i-x_{\ell}$ and $\Delta y_{i\ell}=y_i-y_{\ell}$ denote the difference in position between the $i$-th and $\ell$-th element of the RIS along the $x$ and $y$ axis, respectively and let us approximate Eq.~\eqref{eq:phi_tilde_i_j} as the following
\begin{align}
    \tilde{\phi}_i - \tilde{\phi}_{\ell} & \approx \frac{2\pi}{\lambda}\bigg(\eta_1 - \eta_1 \frac{\psi_z^2}{2}+\eta_2 \psi_z - \eta_1\frac{\psi_y^2}{2}+\eta_3\psi_y\bigg)\Delta x_{i\ell} \nonumber \\
    \phantom{=} & + \frac{2\pi}{\lambda}\bigg(\eta_2 - \eta_2 \frac{\psi_z^2}{2}-\eta_1\psi_z-\eta_2\frac{\psi_x^2}{2}+\eta_3\psi_x\bigg)\Delta y_{i\ell}
\end{align}
where we have neglected the terms that are $\mathcal{O}(\psi^3)$. Let $f_{\psi_x}(\psi_x)$, $f_{\psi_y}(\psi_y)$, and $f_{\psi_z}(\psi_z)$ denote the pdf of $\psi_x$, $\psi_y$, and $\psi_z$, respectively. Hence, we have that 
\begin{align}
     &\Exp\left[e^{j(\tilde{\phi}_{i}-\tilde{\phi}_{\ell})}\right] \nonumber\\
     &=\overset{+\infty}{\underset{-\infty}{\iiint}}  e^{j(\tilde{\phi}_{i}-\tilde{\phi}_{\ell})}f_{\psi_x}(\psi_x)f_{\psi_y}(\psi_y)f_{\psi_z}(\psi_z)d\psi_xd\psi_yd\psi_z\\
     &=\overset{+\infty}{\underset{-\infty}{\iiint}} e^{j\frac{2\pi}{\lambda}\Big(\eta_1\big( \frac{2-\psi_y^2-\psi_z^2}{2}+\frac{\eta_3}{\eta_1}\psi_y+\frac{\eta_2}{\eta_1} \psi_z \big)\Delta x_{i,\ell}+\eta_2\big(\frac{2-\psi_x^2-\psi_z^2}{2}} \nonumber\\
     &\phantom{=} ^{+\frac{\eta_3}{\eta_2}\psi_x-\frac{\eta_1}{\eta_2}\psi_z\big)\Delta y_{i,\ell}\Big)}f_{\psi_x}(\psi_x) f_{\psi_y}(\psi_y)f_{\psi_z}(\psi_z)d\psi_xd\psi_yd\psi_z \\
     &=e^{j\frac{2\pi}{\lambda}(\eta_1\Delta x_{i,\ell}+\eta_2\Delta y_{i,\ell})}\overset{+\infty}{\underset{-\infty}{\iiint}} e^{j\frac{2\pi}{\lambda}\Big(\eta_1\big( -  \frac{\psi_y^2+\psi_z^2}{2}+\frac{\eta_3}{\eta_1}\psi_y+\frac{\eta_2}{\eta_1} \psi_z \big)}\nonumber\\
     &\phantom{=}^{\times\Delta x_{i,\ell} + \eta_2\big( - \frac{\psi_x^2+\psi_z^2}{2}+\frac{\eta_3}{\eta_2}\psi_x-\frac{\eta_1}{\eta_2}\psi_z\big)\Delta y_{i,\ell}\Big)}f_{\psi_x}(\psi_x)f_{\psi_y}(\psi_y)\nonumber\\
     &\phantom{=}\times f_{\psi_z}(\psi_z)d\psi_x d\psi_yd\psi_z.\label{eq:integral}
\end{align}

The integral in Eq.~\eqref{eq:integral} can be seen as the product of three integrals in the form
\begin{align}
    &\int^{\infty}_{-\infty} \frac{1}{\sqrt{2\pi \sigma^2_{\psi}}}e^{-\frac{1}{2}\frac{\psi^2}{\sigma^2_{\psi}}}e^{j\left( - a_{i\ell} \frac{\psi^2}{2}+b_{i\ell} \psi\right)}d\psi\label{eq:integral2_1}\nonumber\\
    &=\frac{1}{\sqrt{2\pi \sigma^2_{\psi}}}\int^{\infty}_{-\infty} e^{-\frac{1}{2}\frac{\psi^2}{\sigma^2_{\psi}} -j a_{i\ell} \frac{\psi^2}{2}+ j b_{i\ell} \psi}d\psi\\
    & =\frac{1}{\sqrt{2\pi \sigma_{\psi}^2}}\int^{\infty}_{-\infty}e^{-\left(\frac{1+j a_{i\ell} \sigma_{\psi}^2}{2\sigma^2_{\psi}}\right)\psi^2 + i b_{i\ell} \psi}d\psi
\end{align}
where the coefficients $a_{i\ell}$ and $b_{i\ell}$ are alternatively expressed as the following
\begin{alignat}{2}
    &a^{(x)}_{i,\ell} && =\frac{2\pi}{\lambda} \eta_2 \Delta y_{i,\ell}\\
    &a^{(y)}_{i,\ell} && =\frac{2\pi}{\lambda} \eta_1 \Delta x_{i,\ell}\\
    &a^{(z)}_{i,\ell} && = \frac{2\pi}{\lambda} \big(\eta_1 \Delta x_{i,\ell}+\eta_2 \Delta y_{i,\ell}\big)
\end{alignat}
and
\begin{alignat}{2}
   &b^{(x)}_{i,\ell} && =\frac{2\pi}{\lambda} \eta_3 \Delta y_{i,\ell}\\
   &b^{(y)}_{i,\ell} && =\frac{2\pi}{\lambda} \eta_2 \Delta x_{i,\ell}\\
   &b^{(z)}_{i,\ell} && = \frac{2\pi}{\lambda} \big(\eta_2 \Delta x_{i,\ell}-\eta_1 \Delta y_{i,\ell}\big),
\end{alignat}
respectively. Let $c_{i,\ell} = \frac{1+j a_{i,\ell} \sigma^2_{\psi}}{2\sigma^2_{\psi}}$ such that
\begin{align}
    &\int^{\infty}_{-\infty} \frac{1}{\sqrt{2\pi \sigma^2_{\psi}}}e^{-\frac{1}{2}\frac{\psi^2}{\sigma^2_{\psi}}}e^{j\left( - a_{i,\ell} \frac{\psi^2}{2}+b_{i,\ell} \psi\right)}d\psi\nonumber\\
    & =\frac{1}{\sqrt{2\pi \sigma^2_{\psi}}}\int^{\infty}_{-\infty}e^{-c_{i,\ell}\psi^2 + j b_{i,\ell} \psi }d\psi\\
    & =\frac{1}{\sqrt{2\pi \sigma^2_{\psi}}}\int^{\infty}_{-\infty}e^{-c_{i,\ell}\left(\psi^2 - \frac{j b_{i,\ell}}{c_{i,\ell}} \psi \right)}d\psi\\
    & =\frac{1}{\sqrt{2\pi \sigma^2_{\psi}}}\int^{\infty}_{-\infty}e^{-c_{i,\ell}\left[\left(\psi - \frac{j b_{i,\ell}}{2c_{i,\ell}} \right)^2 - \left(\frac{j b_{i,\ell}}{2c_{i,\ell}}\right)^2 \right]}d\psi\\
    &= \frac{1}{\sqrt{2\pi \sigma^2_{\psi}}}\int^{\infty}_{-\infty}e^{-c_{i,\ell}\left(\psi - \frac{j b_{i,\ell}}{2c_{i,\ell}} \right)^2 - \frac{b^2}{4c} }d\psi\\
    &= \frac{e^{- \frac{b^2_{i,\ell}}{4c_{i,\ell}}}}{\sqrt{2\pi \sigma^2_{\psi}}}\int^{\infty}_{-\infty}e^{-c_{i,\ell}\left(\psi - \frac{j b_{i,\ell}}{2c_{i,\ell}} \right)^2 }d\psi\nonumber\\
\end{align}
Lastly, by noting that $\mathrm{Re}(c_{i,\ell})>0$ and by operating the following change of variable
\begin{alignat}{2}
   & y_{i,\ell} &&= \psi - \frac{j b_{i,\ell}}{2c_{i,\ell}};\\
    & \frac{dy_{i,\ell}}{d\psi}&&=1;\\
    &d\psi&&=dy_{i,\ell};
\end{alignat}
we have that
\begin{align}
    & \frac{e^{- \frac{b^2_{i,\ell}}{4c_{i,\ell}}}}{\sqrt{2\pi \sigma^2_{\psi}}}\int^{\infty}_{-\infty}e^{-c_{i,\ell}\left(\psi - \frac{j b_{i,\ell}}{2c_{i,\ell}} \right)^2 }d\psi  \nonumber\\
    & =  \frac{e^{- \frac{b_{i,\ell}^2}{4c_{i,\ell}}}}{\sqrt{2\pi \sigma^2_{\psi}}}\int^{\infty}_{-\infty}e^{-c_{i,\ell} y_{i,\ell}^2 }dy_{i,\ell}\label{eq:gauss_integral} \\
      & = \frac{e^{- \frac{b^2_{i,\ell}}{4c_{i,\ell}}}}{\sqrt{2\pi \sigma^2_{\psi}}}\sqrt{\frac{\pi}{c_{i,\ell}}}  \label{eq:gauss_integral_solved}\\
      &=\frac{e^{- \frac{b^2_{i,\ell}}{4\left(\frac{1+j a_{i,\ell} \sigma^2_{\psi}}{2\sigma^2}\right)}}}{\sqrt{2\pi \sigma^2_{\psi}}}\sqrt{\frac{2\pi\sigma^2_{\psi}}{1+j a_{i,\ell}\sigma^2_{\psi}}}\\
      &=\frac{1}{\sqrt{1+j a_{i,\ell}\sigma^2_{\psi}}}e^{- \frac{b^2_{i,\ell}\sigma^2_{\psi}}{2(1+j a_{i,\ell}\sigma^2_{\psi})}}
\end{align}
where Eq.~\eqref{eq:gauss_integral_solved} is found by solving the Gaussian integral in Eq.~\eqref{eq:gauss_integral}. Hence, we have that each element of the matrix $\widetilde{\vv{A}}(\vv{q},\vv{w})$ is approximated as the following
\begin{align}
    [\widetilde{\vv{A}}(\vv{q},\vv{w})]_{i,\ell} &\approx  e^{j\frac{2\pi}{\lambda}\left(\eta_1\Delta x_{i,\ell}+\eta_2\Delta y_{i,\ell}\right)}\frac{e^{- \frac{\big(b^{(x)}_{i,\ell}\big)^2\sigma_x^2}{2\Big(1+j \big(a^{(x)}_{i,\ell}\big)^2\sigma_x^2\Big)}}}{\sqrt{1+j \big(a^{(x)}_{i,\ell}\big)^2\sigma_x^2}}\nonumber\\
    &\phantom{=}\times \frac{e^{- \frac{\big(b^{(y)}_{i,\ell}\big)^2\sigma_y^2}{2\Big(1+j \big(a^{(y)}_{i,\ell}\big)^2\sigma_y^2\Big)}}}{\sqrt{1+j \big(a^{(y)}_{i,\ell}\big)^2\sigma_y^2}}\frac{e^{- \frac{\big(b^{(z)}_{i,\ell}\big)^2\sigma_z^2}{2\Big(1+j\big(a^{(z)}_{i,\ell}\big)^2\sigma_z^2\Big)}}}{\sqrt{1+j \big(a^{(z)}_{i,\ell}\big)^2\sigma_z^2}}.
\end{align}
\end{appendix}

\addcontentsline{toc}{chapter}{References}
\bibliographystyle{IEEEtran}
\bibliography{refs}

\end{document}